\documentclass{jfm}
\usepackage{graphicx}
\usepackage{epstopdf}
\usepackage{epsfig}

\usepackage{soul}
\usepackage{natbib}
\usepackage{bm}
\usepackage{amsmath}
\usepackage{booktabs} %
\usepackage{array} %
\usepackage{paralist} %
\usepackage{verbatim} %
\usepackage{subcaption}

\newcommand{\bh}[1]{ \bm{\hat{#1}} }

\renewcommand{\bar}[1]{\overline{\vphantom{\big(} #1}}
\newcommand{\tr}[1]{ #1^\textrm{T}}
\newcommand{\trace}[1]{\textrm{tr}\left(#1\right)}
\newcommand{\Pe}{\textrm{Pe}}
\newcommand{\Sh}{\textrm{Sh}}
\renewcommand{\Re}{\textrm{Re}}
\newcommand{\iu}{\textrm{i}}

\shorttitle{Mass transfer to freely suspended particles}
\shortauthor{J. M. Lawson}

\title{Mass transfer to freely suspended particles at high P{\'e}clet number}

\author{John M. Lawson\aff{1}
  \corresp{\email{j.m.lawson@soton.ac.uk}}}

\affiliation{\aff{1}Aerodynamics and Flight Mechanics Research Group, University of Southampton, SO17 1BJ, UK}

\begin{document}

\maketitle

\begin{abstract}

In a theoretical analysis, we generalise well known asymptotic results to obtain expressions for the rate of transfer of material from the surface of an arbitrary, rigid particle suspended in an open pathline flow at large P{\'e}clet number, $\Pe$.
The flow may be steady or periodic in time.
We apply this result to numerically evaluate expressions for the surface flux to a freely suspended, axisymmetric ellipsoid (spheroid) in Stokes flow driven by a steady linear shear.
We complement these analytical predictions with numerical simulations conducted over a range of $\Pe = 10^1 - 10^4$ and confirm good agreement at large P{\'e}clet number.
Our results allow us to examine the influence of particle shape upon the surface flux for a broad class of flows.
When the background flow is irrotational, the surface flux is steady and is prescribed by three parameters only: the P{\'e}clet number, the particle aspect ratio and the strain topology.
We observe that slender prolate spheroids tend to experience a higher surface flux compared to oblate spheroids with equivalent surface area.
For rotational flows, particles may begin to spin or tumble, which may suppress or augment the convective transfer due to a realignment of the particle with respect to the strain field.

\end{abstract}


\section{Introduction}

When small, rigid particles are immersed in a fluid, material (e.g. a solute) may be transferred away from the surface by convection and diffusion.
We shall refer to this process as mass transfer, although an analogous problem exists for heat transfer.
The engineered and natural world are replete with examples: planktonic osmotrophs absorb dissolved nutrients \citep{KarpBoss1996,Guasto2012,Pahlow1997}, bacterial hosts encounter viruses \citep{Guasto2012}, crystals are grown in agitated suspension to produce pharmaceuticals and industrial products \citep{HandbookCrystallisation2002} and microplastics leach and sorb pollutants in ocean waters \citep{Law2017,Suhrhoff2016,Seidensticker2017}.
The examples cited above have in common that the solid particle exchanging material is rarely ever spherical, is usually small compared to the flow features in which it is embedded, and the material diffuses slowly such that convection is the dominant mechanism of mass transfer.
The question then naturally follows: what is the rate of mass transfer from the surface?

This question belongs to a general set of classical problems which have been well studied (see e.g. works summarised by \citet{Clift1978,Leal2012,Michaelides2003}).
The answer depends upon the competing effects of convection and diffusion, as parametrised by the P{\'e}clet number, $\Pe$.
When $\Pe \ll 1$, conduction dominates inside an inner region near the particle and the effects of shape are readily accounted for \citep{Batchelor1979,Acrivos1980,Feng1997,Pozrikidis1997}.
When $\Pe \gg 1$, convection dominates the mass transfer process and the surface flux of solute is determined by the particle geometry and the nature of the relative flow field, parametrised by the Reynolds number $\Re$.
In closed streamline flows, the solute simply recirculates around the particle and the surface flux approaches a constant as $\Pe \rightarrow \infty$ \citep{FrankelAcrivos1968,Poe1976}.
When the particle is surrounded by a region of open streamlines (or pathlines), the solution to the convection-diffusion problem takes the form of a thin concentration boundary layer around the particle.
For this case, \citet{Acrivos1960} and later \citet{GoddardAcrivos1965} introduced asymptotic methods to compute the mass transfer rate, which scales as $\Pe^{1/3}$.
The flow and the geometry of the problem then determine the scaling coefficient, which prescribes the surface flux.

For sufficiently small particles, $\Re \ll 1$ and the surrounding relative flow field may be well approximated by a Stokes flow consisting of a background uniform flow or linear shear, plus a perturbation owing to the presence of the particle.
The available analytical results in the high P{\'e}clet number, low Reynolds number regime are generally limited to simplified flows or geometries.
A general solution to this problem would have great utility, because solid particles are often neither spherical nor subject to motions as simple as uniform or axisymmetric flows \citep{Leal2012}.
Specific analytical results are available for spheres in uniform flow \citep{Acrivos1960,AcrivosTaylor1962} or arbitrary linear shear \citep{Gupalo1972,Poe1976,Batchelor1979} and axisymmetric bodies in uniform flow \citep{Sehlin1969,Gupalo1976,Leal2012,Dehdashti2020}.
To our knowledge, equivalent results are not available for arbitrary bodies with high P{\'e}clet numbers in an arbitrary linear shear.
Experimental and numerical studies have focused on the uniform flow case typically with $\Pe = O(\Re)$ \citep{Masliyah1972,Clift1978,Sparrow2004,Ke2018,Kishore2011,Ma2020}; relatively few studies have examined the high P{\'e}clet number, low Reynolds number regime numerically \citep{KarpBoss1996,Pahlow1997} and the results available for linear shear flows are limited to a handful of cases.
Therefore, additional empirical data are needed for the general case of arbitrary linear shear to support a generalisation of asymptotic results.

For freely suspended particles, an additional complication arises which affects the mass transfer rate.
In the absence of body forces, such as the case of neutrally buoyant particles, the drag and consequently the slip velocity vanish, such that convection is provided by the linear shear alone.
However, the fluid may exert a couple upon the particle \citep{Jeffery1922}, which in the absence of a restoring torque will result in the steady precession or rotation of the particle about its axis.
As a result, the flow field around the particle and resultant surface flux is in general unsteady \citep{Pahlow1997}.
However, for spherical particles, \citet{Batchelor1979} and \citet{Batchelor1980} argued that the unsteady contribution to the scalar flux may be neglected at large P{\'e}clet number, so that the average flow field perceived by the body determines the average mass transfer rate.

In this article, we extend the work of \citet{GoddardAcrivos1965} and \citet{Batchelor1979} to consider the mass transfer rate due to convection and diffusion from an arbitrary body in an unsteady, open pathline flow at high P{\'e}clet number.
We then apply this analysis to determine the mass transfer rate from a neutrally buoyant spheroid freely suspended in an arbitrary linear shear.
We obtain asymptotic expressions for the transfer rate in the resultant average flow field and compare these to the results of numerical simulations, which provide a quantitative test of the accuracy of the asymptotic approximation.
Our results allow us to examine the role of particle shape in the mass transfer process for a very broad class of flows and open up new possibilities in the numerical simulation of mass transfer in particle suspensions.

The paper is structured as follows.
In \S\ref{sec:general-theory}, we review the theoretical background of the problem and derive a general form for the mass transfer coefficient for an arbitrary particle in an unsteady, time periodic flow.
In \S\ref{sec:ellipsoid-theory}, we apply this general expression to the case of a freely suspended spheroid in Stokes flow.
In \S\ref{sec:sim}, we introduce a numerical test of the results given in \S\ref{sec:ellipsoid-theory} and discuss the influence of particle shape upon the surface flux.
We present a summary of our results and future perspectives in \S\ref{sec:conclusions}.

\section{The steady flux at large P{\'e}clet number}
\label{sec:general-theory}

In this section, we shall extend the analyses of \citet{GoddardAcrivos1965} and \citet{Batchelor1979} to derive a general expression for the average solute flux from the surface of a particle of arbitrary shape in a steady, open-streamline flow.
We begin by introducing the governing equations in dimensionless form, then examine the case of steady flow.
Finally, we generalise the result to unsteady flow.

\subsection{Governing equations}
\label{sec:governing-equation}

The mass transport from the surface of the particle is governed by the convection-diffusion equation.
This may be written in dimensionless form as
\begin{equation}
\label{eqn:scalar-transport}
\frac{\partial \theta}{\partial t} + \bm{u}\cdot\nabla \theta = \frac{1}{\Pe}\nabla^2 \theta
\end{equation}
where $\bm{u}(\bm{x},t)$ is the velocity field and $\theta(\bm{x},t)$ is the concentration of the solute \citep{Leal2012}.
The characteristic lengthscale is $r$, the linear dimension of the particle, so that the spatial coordinate is non-dimensionalised as $\bm{x} = \bm{x}^*/r$.
The characteristic timescale is prescribed by the shear rate $E^*$, so that time is non-dimensionalised as $t=t^*E^*$.
The P{\'e}clet number is defined $\Pe \equiv r^2 E^* / \kappa$, where $\kappa$ is the diffusion coefficient of the solute.
Our convention is to write dimensional quantities with a superscript $^*$ unless otherwise stated.

We shall adopt a frame of reference moving with the particle, such that the boundary of the particle $S_p$ is stationary. 
We impose the boundary condition
\begin{equation}
\label{eqn:boundary-condition}
\begin{matrix}
\theta = 1 \text{ and } \bm{u} = 0 	&\text{for } &\bm{x} \in S_p \\
\theta \rightarrow 0 				&\text{as } &|\bm{x}| \rightarrow \infty
\end{matrix}
\end{equation}
so that the concentration of solute at the particle surface is uniform.
This boundary condition corresponds to non-dimensionalising the concentration field $C(\bm{x}^*,t^*)$ as $\theta = (C-C_0)/(C_1-C_0)$, where $C_1$ and $C_0$ are the concentration of solute at the surface and infinity in physical units.

The non-dimensional measure of the surface flux $Q$ is the Sherwood number 
\begin{equation}
\label{eqn:sherwood}
\Sh = \frac{Q}{4\pi r\kappa(C_1 - C_0)}
= -\frac{1}{4\pi} \iint_{S_p} \nabla \theta \cdot \mathrm{d}\mathbf{S}.
\end{equation}
The denominator in (\ref{eqn:sherwood}) corresponds to the steady state flux from a sphere of radius $r$ in the absence of flow.
A convenient choice for the lengthscale $r$ is $\sqrt{A/4\pi}$, where $A$ is the surface area of the particle.
The Sherwood number is therefore the factor by which convection enhances the mass transfer relative to the diffusive flux from a spherical particle with the same surface area. 

\subsection{General solution of the surface flux in steady flow}
\label{sec:general-solution} 

In steady flow, the scalar transport equation reduces to
\begin{equation}
\label{eqn:steady-scalar}
\bm{u}\cdot \nabla \theta = \frac{1}{\Pe} \nabla^2 \theta
\end{equation}
We seek an asymptotic solution for (\ref{eqn:steady-scalar}) subject to (\ref{eqn:boundary-condition}) at large P{\'e}clet number.

At large P{\'e}clet number, the concentration of solute is small almost everywhere far from the particle, apart from a thin wake, along which material from the surface is swept.
Near the particle, there is a concentration boundary layer whose thickness is $O(\delta)$, across which there is a sharp jump in concentration.
For the boundary layer approximation to hold, the pathlines near the surface of the particle must originate and terminate at infinity, such that the surface of the particle is exposed to a constant stream of fresh fluid.
Recirculating regions (closed pathlines), which can occur in some geometries, are not permitted.

\begin{figure}
\centering
\includegraphics[width=0.5\textwidth,trim=0cm 2.5cm 0cm 1.5cm,clip]{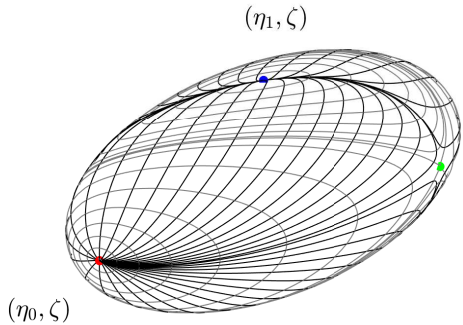}
\caption{Illustration of the curvilinear coordinate system $(\xi,\eta,\zeta)$ defined on the surface ($\xi=0$) of a spheroid in an arbitrary linear shear. Thick black lines show $\eta$-coordinate pathlines ($\xi=0,\zeta=\textrm{const.}$), which are tangent to the local viscous shear stress on the surface. Thin grey lines are $\zeta$-coordinate pathlines ($\xi = 0, \eta=\textrm{const.}$). Three fixed points of the surface streamlines are shown: red (source), green (saddle) and blue (sink).}
\label{fig:coordinates}
\end{figure}

We shall construct a general curvilinear coordinate system $(\xi, \eta, \zeta)$ to describe the flow near the surface in terms of the distance from the surface and the direction of the flow near the surface.
This coordinate system is not necessarily orthogonal and we will find it useful to describe it in terms of the covariant coordinate vectors 
\begin{equation}
\label{eqn:covariant}
\bm{h}_\xi  = \frac{\partial \bm{x}}{\partial \xi}
~,~
\bm{h}_\eta = \frac{\partial \bm{x}}{\partial \eta}
~,~
\bm{h}_\zeta= \frac{\partial \bm{x}}{\partial \zeta}
\end{equation}
and their pathlines, which we have illustrated in figure \ref{fig:coordinates}.
We note that the adoption of a general curvilinear coordinate system, rather than an orthogonal one, is crucial in the generalisation to an arbitrary flow or body.

The coordinate $\xi$ is defined as the normal distance from the particle surface, so that at the particle surface, $\bm{h}_\xi$ is a unit vector normal to the surface.
The $\eta$ and $\zeta$ coordinates lie tangent to the surface.
We shall define the $\eta$ coordinate so that at a small distance $\xi$ above the surface, the component of fluid velocity which is tangent to the surface is parallel to $\bm{h}_\eta$.
The curves along the $\eta$ direction are therefore ``surface streamlines'' which are tangent to the direction of the surface velocity gradient $\bm{w} = \partial \bm{u}/\partial \xi|_{\xi=0}$.

We can therefore express the velocity near the surface as 
\begin{equation}
\begin{aligned}
\bm{u}
&= u^\xi \bm{h}_\xi + u^\eta\bm{h}_\eta + u^\zeta\bm{h}_\zeta \\
&= \bm{w}\xi + O(\xi^2).
\end{aligned}
\end{equation}
so that near the particle surface, the velocity components are of the form
\begin{equation}
\label{eqn:u-taylor-series}
u^\xi = G(\eta,\zeta)\xi^2 + O(\xi^3), u^\eta = \xi F(\eta,\zeta) + O(\xi^2), u_\zeta = O(\xi^2)
\end{equation}
By definition, the velocity at the surface is zero; the tangential component $u^\eta$ is linear in $\xi$ to leading order, whereas the surface-normal component is quadratic \citep{Leal2012}.
The function $F(\eta, \zeta)$ is therefore obtained in terms of the velocity gradient at the surface of the particle as
\begin{equation}
\label{eqn:F}
F = \bm{w} \cdot \nabla \eta
\end{equation}
and is related to $G(\eta,\zeta)$ through the continuity equation.

To obtain the surface normal velocity component $u^\xi$, we express the continuity equation as \citep{Grinfeld2013} 
\begin{equation}
\label{eqn:continuity}
\rho\nabla_i u^i = 
  \frac{\partial (\rho u^\xi)}{\partial \xi}
+ \frac{\partial (\rho u^\eta)}{\partial \eta}
= 0
\end{equation}
where $\rho = \rho(\eta, \zeta) \equiv \sqrt{\det{g}}$ and $\det{g}$ is the determinant of the metric tensor $g_{ij} \equiv \bm{h}_i\cdot\bm{h}_j$.
Since $g_{\xi\xi} = 1$, $g_{\xi\eta} = g_{\xi\zeta} = 0$ by definition, the physical interpretation of $\rho$ is a local area density of the surface streamlines, such that $\delta A = \rho \delta \eta \delta \zeta$ is the area covered by a small region on the surface measuring $\delta \eta$ along a surface streamline and $\delta \zeta$ across adjacent streamlines.
We can then define a streamfunction $\psi$ as
\begin{equation}
\label{eqn:streamfunction}
u^\xi = -\frac{1}{\rho} \frac{\partial \psi}{\partial \eta}
~,~
u^\eta = \frac{1}{\rho} \frac{\partial \psi}{\partial \xi}
\end{equation}
so that
\begin{equation}
\psi = \frac{1}{2}\xi^2 \rho F %
\end{equation}
is a solution which satisfies (\ref{eqn:u-taylor-series}) and the continuity equation (\ref{eqn:continuity}) to leading order in $\xi$.

Writing (\ref{eqn:steady-scalar}) in these curvilinear coordinates, our solution now proceeds analogously to the works of \citet{GoddardAcrivos1965} and \citet{Batchelor1979}.
We shall formalise Batchelor's analysis here in our new curvilinear coordinate system for the purpose of exposition.
To leading order in $\xi$, the convection term is
\begin{equation}
\label{eqn:convection-approximation}
u^i \nabla_i \theta 
= -\frac{1}{\rho}\frac{\partial \psi}{\partial \eta}\frac{\partial \theta}{\partial \xi}
  +\frac{1}{\rho}\frac{\partial \psi}{\partial \xi}\frac{\partial \theta}{\partial \eta}
  + O(\xi^2)
\end{equation}
whereas the diffusion term may be written
\begin{equation}
\label{eqn:laplacian-approximation}
\nabla^2 \theta
=
\frac{1}{\rho}
\left[
 \frac{\partial}{\partial \xi^i}\left(\rho g^{ij} \frac{\partial \theta}{\partial \xi^j}\right)
\right]
=
\frac{1}{\rho}
 \frac{\partial}{\partial \xi}\left(\rho \frac{\partial \theta}{\partial \xi}\right)
+ \ldots
\end{equation}
where $g^{ij}$ is the inverse metric tensor and $\xi^i$ indexes the coordinates $(\xi,\eta,\zeta)$.
The terms omitted in the expansion of (\ref{eqn:laplacian-approximation}) do not involve any gradients in $\xi$.
By neglecting them, we assume that the diffusive flux of material across the surface is small in comparison to that normal to the surface.
This essentially requires that the particle surface be smooth and contain no regions of extreme curvature where this assumption may break down.

We now rescale our coordinate system and streamfunction so that the surface-normal concentration gradient is $O(1)$.
Rescaling $\xi$ by the boundary layer thickness $\delta$, we write
\begin{equation}
\label{eqn:rescaling}
\Xi = \frac{\xi r}{\delta} = \xi\Pe^m 
~,~
\Psi = \frac{1}{2}\Xi^2 \rho F = \Pe^{2m} \psi
\end{equation}
where we suppose $\delta/r$ scales as $\Pe^{-m}$.
We can rewrite (\ref{eqn:steady-scalar}) using (\ref{eqn:convection-approximation}) and (\ref{eqn:laplacian-approximation})
\begin{equation}
\label{eqn:transport-rescaled}
-\frac{\partial \Psi}{\partial \eta}\frac{\partial \theta}{\partial \Xi}
+\frac{\partial \Psi}{\partial \Xi}\frac{\partial \theta}{\partial \eta}
+O(\Pe^{-m})
=
\Pe^{3m-1} \frac{\partial}{\partial \Xi}\left(\rho \frac{\partial \theta}{\partial \Xi}\right)
\end{equation}
The first two terms in (\ref{eqn:transport-rescaled}) are $O(1)$ and thus $m=1/3$, as expected.

We can now solve (\ref{eqn:transport-rescaled}) with the well-known von-Mises transformation.
Adopting the change of variables $(\Xi,\eta,\zeta) \rightarrow (\Psi,\eta,\zeta)$, we have
\begin{equation}
\label{eqn:c-psi}
\frac{\partial \theta}{\partial \eta}
= \frac{\partial}{\partial \Psi}\left(
\rho \frac{\partial \theta}{\partial \Psi} \frac{\partial \Psi}{\partial \Xi}
\right)
= \frac{\partial}{\partial \Psi}\left(
\frac{\partial \theta}{\partial \Psi}(2\rho^3 F\Psi)^{1/2}
\right).
\end{equation}
Equation (\ref{eqn:c-psi}) now admits a self-similar solution of the form $\theta = \theta(\chi), \chi = \Psi^{1/2}/\tau^{1/3}$, where the functions $\theta(\chi)$ and $\tau(\eta)$ must satisfy
\begin{equation}
\label{eqn:theta}
\frac{\mathrm{d}^2 \theta}{\mathrm{d} \chi^2} 
+ \frac{4}{3}\chi^2 \frac{\mathrm{d} \theta}{\mathrm{d} \chi} = 0
\end{equation}
for
\begin{equation}
\label{eqn:tau}
\frac{\mathrm{d}\tau}{\mathrm{d}\eta} = (2\rho^3 F)^{1/2}.
\end{equation}
The solution of (\ref{eqn:theta}) satisfying the imposed boundary conditions (\ref{eqn:boundary-condition}) of $\theta(0) = 1$ and $\theta = 0$ as $\chi \rightarrow \infty$ is
\begin{equation}
\label{eqn:theta-solution} 
\theta(\chi) = \frac{\Gamma(\frac{1}{3}, \frac{4}{9} \chi^3)}{\Gamma(\frac{1}{3},0)} %
\end{equation}
where $\Gamma$ is the incomplete gamma function.
We can obtain $\tau$ by integrating (\ref{eqn:tau}) along the surface streamlines, which must begin and terminate at critical points $F=0$ on the surface, since the surface shear stress is continuous. 
We will therefore choose the constant of integration so that
\begin{equation}
\tau(\eta) = \int_{\eta_0}^{\eta} (2\rho^3 F)^{1/2} \mathrm{d}\eta
\end{equation}
has $\tau(\eta_0) = 0$ at the beginning of the surface streamline $\eta=\eta_0$ and $\tau(\eta_1) = \tau_1$ at the end $\eta = \eta_1$.

At the surface, the local flux of material per unit area is
\begin{equation}
\begin{aligned} 
-\frac{\partial \theta}{\partial \xi}\Big|_{\xi = 0}
&= -\Pe^{1/3} \frac{\partial \theta}{\partial \Xi}\Big|_{\Xi=0} 
= -\Pe^{1/3} \frac{\partial \theta}{\partial \chi}\Big|_{\chi=0} 
\cdot\frac{\partial \chi}{\partial \Xi}\Big|_{\Xi=0} \\
&= 
\Pe^{1/3}
\frac{12^{1/3}}{\Gamma(\frac{1}{3})}
\cdot
\frac{\left(\frac{1}{2}\rho F\right)^{1/2}}{\tau^{1/3}}
\end{aligned}
\end{equation}
The Sherwood number (\ref{eqn:sherwood}) is therefore given by
\begin{equation}
\begin{aligned}
\Sh 
&= -\frac{1}{4\pi} 
\iint_{S_p}  \frac{\partial \theta}{\partial \xi}\Big|_{\xi=0} 
\rho\mathrm{d}\eta \mathrm{d}\zeta \\
&= \frac{12^{1/3} \Pe^{1/3}}{8\pi \Gamma(1/3)} 
\iint_{S_p} \frac{\left(2 \rho^3 F\right)^{1/2}}{\tau^{1/3}}
\mathrm{d}\eta \mathrm{d}\zeta 
\end{aligned}
\end{equation}
which we can integrate by parts by recognising that
\begin{equation}
\int_{\eta_0}^{\eta_1} \frac{\left(2 \rho^3 F\right)^{1/2}}{\tau^{1/3}} \mathrm{d}\tau
= 
\int_{\eta_0}^{\eta_1} \frac{\mathrm{d}\tau}{\mathrm{d}\eta} \tau^{-1/3} \mathrm{d}\eta
= 
\frac{3}{2} \tau_1^{2/3}
\end{equation}
and thus
\begin{equation}
\label{eqn:sherwood-solution}
\Sh 
= \frac{0.808 \Pe^{1/3}}{4\pi} \int \left(
\int_{\eta_0}^{\eta_1}
\rho^{3/2} F^{1/2} \mathrm{d}\eta
\right)^{2/3}\mathrm{d}\zeta  
+ O(1)
\end{equation}
As expected, the limiting dimensionless flux scales as $\alpha \Pe^{1/3}$, with a prefactor $\alpha$ which can now be explicitly computed in terms of the shape of the body and the surface velocity gradient.

The main point of (\ref{eqn:sherwood-solution}) is that we have generalised the results of \citet{GoddardAcrivos1965} and \citep{Batchelor1979} to any distribution of surface streamlines on an arbitrary body, not just those which result in an orthogonal coordinate system.
The caveat remains that the pathlines around the body should be open for the boundary layer approximation to hold and the boundary should be smooth and contain no regions of extreme curvature.
Furthermore, we have a natural way of numerically constructing the local area density $\rho$, as shown graphically in figure \ref{fig:coordinates}.
By integrating (\ref{eqn:covariant}) with respect to $\eta$ to generate a set of surface streamlines over the body, we obtain lines of constant $\zeta$ at intervals of varying $\eta$.
With a sufficiently fine discretisation, we can numerically approximate the local area density $\rho$ and evaluate the coefficient in (\ref{eqn:sherwood-solution}) numerically.
We shall demonstrate this procedure later in \S\ref{sec:ellipsoid-theory}.

\subsection{Unsteady solution at large P{\'e}clet number}

We now seek to generalise our steady solution to an unsteady, periodic flow.
Our argument is analogous to that proposed by \citet{Batchelor1980}, who examined the scalar flux to spherical particles in turbulent flow.

We shall consider the case where the motion is periodic with period $T = E^*T^*$.
We seek the time average Sherwood number $\bar{\Sh}$ at the surface, which depends upon the time average concentration field $\bar{\theta}$.
The time average over a period $T$ is defined simply as
\begin{equation}
\bar{\theta}(\bm{x}) = \frac{1}{T} \int_{0}^{T} \theta(\bm{x},t) \mathrm{d}t.
\end{equation}
Applying this averaging procedure to (\ref{eqn:scalar-transport}), we obtain
\begin{equation}
\label{eqn:mean-scalar-transport}
\bar{\bm{u}} \cdot \nabla\bar{\theta} + \bar{\bm{u}' \cdot \nabla \theta'} = \frac{1}{\Pe} \nabla^2 \bar{\theta}
\end{equation}
where we have decomposed the velocity field $\bm{u} = \bar{\bm{u}} + \bm{u}'$ and concentration field $\theta =\bar{\theta}+\theta'$ into their time respective average and fluctuating components.
Likewise, the transport equation for the concentration fluctuation $\theta'(\bm{x}, t)$ is
\begin{equation}
\label{eqn:scalar-fluctuation}
\frac{\partial \theta'}{\partial t}
+ \bm{u}\cdot\nabla \theta' 
- \bar{\bm{u}'\cdot\nabla \theta'}
- \frac{1}{\Pe} \nabla^2 \theta'
=
\bm{u}'\cdot \nabla \bar{\theta}
\end{equation}

We shall argue that the solution of (\ref{eqn:scalar-fluctuation}) satisfying the boundary conditions (\ref{eqn:boundary-condition}) and 
\begin{equation}
\theta' = 0 \text{ at } \xi = 0 \text{ and } \theta' = 0 \text{ as } \xi \rightarrow \infty
\end{equation}
is such that $\theta' \sim \Pe^{-1/3}$ as $\Pe \rightarrow \infty$.
As before, we reason that the solution consists of a slender concentration boundary layer of thickness $\delta/r \sim \Pe^{-1/3}$.
Outside the boundary layer $\xi \gg \delta$, $\bar{\theta} \rightarrow 0$ and $\theta' \rightarrow 0$.
Inside the boundary layer, the amplitude of the concentration fluctuations are $\theta' = O(\Pe^{m'})$ and the jump in the mean concentration across the boundary layer is $O(1)$.
Clearly, $m' \le 0$ since there is no mechanism in (\ref{eqn:scalar-transport}) which can amplify scalar fluctuations.
If $m' < 0$, then scalar fluctuations should decay in amplitude at large $\Pe$, whereas if $m' = 0$ they remain comparable in magnitude to the mean flow.

Let us examine the order of magnitude of the terms in (\ref{eqn:scalar-fluctuation}).
Since scalar fluctuations are $O(\Pe^{m'})$ and occur on a timescale of $O(T)$, the time derivative term scales as
\begin{equation}
\frac{\partial \theta}{\partial t}
\sim \frac{\Pe^{m'}}{T}.
\end{equation}
The convective terms on the left hand side scale as
\begin{equation}
\bm{u}\cdot \nabla \theta'
= u_\xi \frac{\partial \theta'}{\partial \xi} + \ldots
\sim \left(\frac{\delta^2}{r^2}\right) \left(\frac{r}{\delta} \Pe^{m'} \right) \sim \Pe^{m'-1/3}
\end{equation}
whereas convective term on the right hand side scales as
\begin{equation}
\bm{u}' \cdot \nabla \bar{\theta} = 
u_\xi' \frac{\partial \bar{\theta}}{\partial \xi} + \ldots
\sim \left(\frac{\delta^2}{r^2}\right) \left(\frac{r }{\delta}\right) \sim \Pe^{-1/3}.
\end{equation}
Finally, the diffusion term scales as
\begin{equation}
\frac{1}{\Pe}\nabla^2 \theta' 
\approx \frac{1}{\Pe} \frac{\partial^2 \theta'}{\partial \xi^2}
\sim \frac{1}{\Pe} \frac{r^2}{\delta^2} \Pe^{m'}
\sim \Pe^{m' - 1/3}.
\end{equation}
Annotating (\ref{eqn:scalar-fluctuation}) with the order of magnitude of each term, we have
\begin{equation}
\label{eqn:scalar-fluctuation-annotated}
\underbrace{\vphantom{\Big[}\frac{\partial \theta'}{\partial t}}_{O(\Pe^{m'}/T)}
\underbrace{\vphantom{\Big[}
+ \bm{u}\cdot\nabla \theta' 
- \bar{\bm{u}'\cdot\nabla \theta'}
- \frac{1}{\Pe} \nabla^2 \theta'
}_{O(\Pe^{m'-1/3})}
=
\underbrace{\vphantom{\Big[}\bm{u}'\cdot \nabla \bar{\theta}}_{O(\Pe^{-1/3})}
\end{equation}
We see that the solution requires $m'=-1/3$, such that the first term on the left hand side dominates and balances the unsteady convection term on the right hand side, provided the dimensionless period $T \ll \Pe^{1/3}$.
Thus, at large P{\'e}clet, we have
\begin{equation}
\frac{\partial \theta'}{\partial t} \approx \bm{u}'\cdot \nabla \bar{\theta} \sim \Pe^{-1/3}
\end{equation}
which is equivalent to equation (3.9) of \citet{Batchelor1980}.
However, we have derived the result on more general grounds.
Physically, this result means that when the timescale of diffusion within the boundary layer becomes large in comparison to the timescale of the unsteadiness, there is insufficient time for diffusion to redistribute scalar fluctuations.
Instead, scalar fluctuations inside the boundary layer occur due to the convection of the mean concentration field by unsteady velocity fluctuations.

Thus, to leading order, the concentration and its time mean are equal and the mean scalar field is well approximated by
\begin{equation}
\bar{\bm{u}}\cdot\nabla\bar{\theta} = \frac{1}{\Pe}\nabla^2 \bar{\theta} + O(\Pe^{-2/3})
\end{equation}
This allows us to apply our general result (\ref{eqn:sherwood-solution}) to unsteady, time periodic flows where the period of motion is sufficiently small.
As before, for the boundary layer approximation to hold, it is required that the surface of the particle be exposed to a continuous stream of fresh fluid.
Therefore, it is required that the pathlines of fluid parcels adjacent to the surface be open, such that the fluid does not simply recirculate around a closed path.

\section{The steady flux for a spheroid}
\label{sec:ellipsoid-theory}

We shall now apply the results of \S\ref{sec:general-solution} to obtain the scaling coefficient $\alpha$ for a spheroid in an arbitrary linear shear.
Although the motion is in general unsteady, we have argued in \S\ref{sec:general-solution} that the average mass transfer rate can be computed for an equivalent spheroid subject to the same average relative flow field.
This flow field can be described by three parameters, but for a broad class of cases, the flow is reduced to an axisymmetric configuration described by a single parameter.
To proceed, we shall analyse the relative motion of the particle in \S\ref{sec:particle-motion} and obtain expressions for the average flow field perceived by the particle.
We introduce an expression for the velocity field near the particle in \S\ref{sec:stokes-flow}, then derive the mass transfer coefficient in the axisymmetric and general cases in \S\ref{sec:axisymmetric-flux} and \S\ref{sec:general-flux} respectively.
Based on these results, we consider the special case of rotation dominated flow in \S\ref{sec:largew} and discuss potential extensions in \S\ref{sec:generalisations}.

\subsection{Motion of a freely suspended spheroid}
\label{sec:particle-motion}

Let us consider the motion of a spheroidal particle in an arbitrary, linear shear flow.
The background linear shear is $\bm{v} = \bm{Gy}$, where $\bm{y}=\bm{y}^*/r$ is the coordinate system of the fixed laboratory frame and $\bm{G} = \bm{E} + \bm{W}$ is an arbitrary, steady velocity gradient in this reference frame.
The velocity gradient is composed of a symmetric strain tensor $\bm{E}=\bm{E}^*/E^*$ and antisymmetric rotation tensor $\bm{W}=\bm{W}^*/E^*$.
We shall choose the characteristic shear rate $E^* = (E_{ij}^*E_{ij}^*)^{1/2}$ so that $E_{ij}E_{ij}=1$.

The spheroid is centred at the origin and its semiaxes $a_i = (a,c,c)$ are spanned by an orthogonal set of unit vectors $\bm{p},\bm{q},\bm{r}$.
Thus, the coordinate $\bm{x}$ in the body frame maps to the laboratory frame as $\bm{y} = \bm{Rx}$, where $\bm{R} = [\bm{p}, \bm{q}, \bm{r}]$.
The unit vector $\bm{p}$ points along the symmetry axis towards the ``pole'' of the spheroid, whilst $\bm{q}$ and $\bm{r}$ point radially outward around the ``equator''. 
The body rotates with angular velocity $\bm{\Omega}=\bm{\Omega}(t)$, such that $\dot{\bm{R}} = [\bm{\Omega}]_\times \bm{R}$, and the velocity field in the body frame is
\begin{equation}
\label{eqn:velocity-body-frame}
\bm{u}(\bm{x},t) = \bm{R}^\textrm{T}\bm{v}(\bm{R}\bm{x},t) - \bm{\Omega}\times(\bm{Rx}).
\end{equation}
Thus, the background velocity gradient $\bm{u}=\bm{A x}$ appears in the body frame as
\begin{equation}
\label{eqn:vgt-body-frame}
\bm{A} = \bm{R}^\textrm{T} (\bm{G} - [\bm{\Omega}]_\times) \bm{R}
\end{equation}
and, in general, varies in time as the particle rotates.

\citet{Jeffery1922} showed that, when the couple upon a spheroid is zero, the solid body rotation rate $\bm{\Omega}$ of the body is given by
\begin{equation}
\bm{\Omega} = \frac{1}{2}\bm{\omega} + \frac{\lambda^2-1}{\lambda^2+1} \bm{p}\times \bm{E}\bm{p},
\end{equation}
where $\omega_i = -\epsilon_{ijk}W_{jk}$ is the background vorticity in the laboratory frame.
Thus, the orientation of the particle $\bm{p}$ then evolves according to Jeffery's equation
\begin{equation}
\label{eqn:pdot}
\begin{aligned}
\dot{\bm{p}} 
&= \bm{W}\bm{p} + 
\frac{\lambda^2-1}{\lambda^2+1}
\left(
\bm{Ep} - \tr{\bm{p}}\bm{Ep}
\right)
\end{aligned}
\end{equation}
where $W_{ij} = \frac{1}{2}\epsilon_{ijk}\omega_k$ is the rate of rotation tensor.

When the background shear is constant in time, the solution to (\ref{eqn:pdot}) can be written in terms of a matrix exponential as
\begin{equation}
\label{eqn:p}
\bm{p}(t) = 
\frac{\breve{\bm{p}}}{|\breve{\bm{p}}|}
~,~
\breve{\bm{p}}(t) = \exp\left(\bm{K}t\right) \bm{p}_0
~,~
\bm{K} = \bm{W} + \frac{\lambda^2-1}{\lambda^2+1}\bm{E}
\end{equation}
subject to the initial condition $\bm{p}=\bm{p}_0$ at $t=0$ \citep{Szeri1993}.

We shall now examine the fixed points and stable attractors of (\ref{eqn:pdot}).
This analysis elaborates upon the previous work by \citet{Bretherton1962}.
Decomposing $\bm{K}$ in terms of its eigenvectors $\bm{e}_i$ and eigenvalues $\lambda_i$, we can write (\ref{eqn:p}) as
\begin{equation}
\label{eqn:p-breve}
\breve{\bm{p}}
=
\sum_i (\bm{p}_0\cdot\bm{e}_i) \exp(\lambda_i t) \bm{e}_i.
\end{equation}
From (\ref{eqn:p-breve}) we see that the fixed points of (\ref{eqn:pdot}) coincide with the eigenvectors $\bm{e}_i$.
The stability of the fixed points depends upon the eigenvalues $\lambda_i$, and because $\sum_i \lambda_i =0$, there is always only one (neutrally) stable fixed point or limit cycle (excepting $\lambda_i=0$).
Thus after a finite time, the motion of the particle will always approach a stable attractor whose nature depends upon the largest eigenvalue(s) of $\bm{K}$.

We can categorise the stable attractors into five different cases: 1a, 1b, 2a, 2b and 3.
These five cases map to four different motions: resting, spinning, 2D (two-dimensional) tumbling and 3D (three-dimensional) tumbling.
These are summarised in table \ref{tbl:spheroid-motion}.
We shall now examine each case.

\begin{table}
\centering
\begin{tabular}{lcccc}
Case & Eigenvalues                                                                     & Condition          & Motion      & $\bar{\bm{A}}$                               \\
1a   & $\lambda_i \in \mathbb{R}, \lambda_1 \le \lambda_2 \le \lambda_3$                                         & $\Omega_3 \ne 0$ & Spinning    & (\ref{eqn:average-gradient-1a}) \\
1b   & ''                                                                              & $\Omega_3 = 0$     & Resting     & (\ref{eqn:average-gradient-1b}) \\
2a   & $\lambda_1 = \lambda_2^\dagger = \sigma + \mathrm{i}\omega, \lambda_3=-2\sigma$ & $\sigma > 0$       & Spinning    & (\ref{eqn:average-gradient-1a}) \\
2b   & ''                                                                              & $\sigma < 0$       & 2D tumbling &                                              \\
3    & ''                                                                              & $\sigma = 0$       & 3D tumbling &                                             
\end{tabular}
\caption{Classification of the motion and time average perceived velocity gradient experienced by a spheroid.}
\label{tbl:spheroid-motion}
\end{table}

In cases 1a and 1b, all the eigenvalues $\lambda_1 \le \lambda_2 \le \lambda_3$ of $\bm{K}$ are real and (\ref{eqn:p}) has three fixed points $\bm{p} = \bm{e}_i$, where $\bm{e}_i$ are the eigenvectors of $\bm{K}$. 
From (\ref{eqn:p-breve}) we see that only the fixed point corresponding to the largest eigenvalue $\lambda_3$ is stable, so for nearly any initial condition, the particle will reorient itself to be parallel to $\bm{e}_3$.
In this configuration, the particle will in general rotate about its axis with angular velocity $\bm{\Omega} = \Omega_3 \bm{e}_3$, where $\Omega_3 = \bm{\omega}\cdot\bm{e}_3/2$ (case 1a).
We call this motion spinning.
However, for particular configurations, $\Omega_3 = 0$ and the particle does not rotate (case 1b).
We call this motion resting.

In cases 2a, 2b and 3, $\bm{K}$ has a pair of complex eigenvalues and the system has one fixed point and a periodic point.
In these cases, we can write the eigenvalues as $\lambda_1 = \lambda_2^\dagger = \sigma + \iu\omega, \lambda_3 = -2\sigma$, where $\dagger$ denotes the complex conjugate.
When $\sigma < 0$, the fixed point is stable and the periodic point is unstable (case 2a, spinning).
Thus the particle aligns with $\bm{e}_3$ as in cases 1a and 1b, but unlike case 1b, the particle must rotate about this axis because the angular velocity $\bm{\Omega} = \Omega_3 \bm{e}_3$ is bounded $\Omega_3^2 \ge \omega^2 > 0$.
When $\sigma > 0$, the fixed point is unstable and the periodic point is stable (case 2b).
Thus the particle will evolve towards a planar limit cycle oscillation described by
\begin{equation}
\label{eqn:p-case2b}
\breve{\bm{p}} = \bm{a}\cos(\omega t) + \bm{b}\sin(\omega t)
\end{equation}
where $\bm{e}_1 = \bm{e}_2^\dagger = \bm{a}+\iu\bm{b}$.
We call this motion 2D tumbling.
In case 3, $\sigma = 0$ and the motion follows three-dimensional Jeffery orbits \citep{Jeffery1922}, where $\bm{p}$ precesses around the axis $\bm{e}_3$ with period $2\pi/\omega$.
We call this motion 3D tumbling.
\footnote[2]{Jeffery orbits are in general 3D, but for some initial conditions, the motion can be reduced to spinning or 2D tumbling.}

The solution of (\ref{eqn:mean-scalar-transport}) depends upon the average flow field in the particle frame.
The instantaneous flow field around the particle consists of a superposition of the background gradient $\bm{Ax}$ and a perturbation owing to the presence of the particle.
Since this perturbation is linear in $\bm{A}$, the time average flow field in the particle frame is determined by the average velocity gradient perceived by the particle.
We shall now calculate the average velocity gradient perceived by the particle in the general limiting cases identified above.

\begin{figure}
\centering
\begin{subfigure}[b]{0.32\linewidth}
	\includegraphics[clip,width=\linewidth]{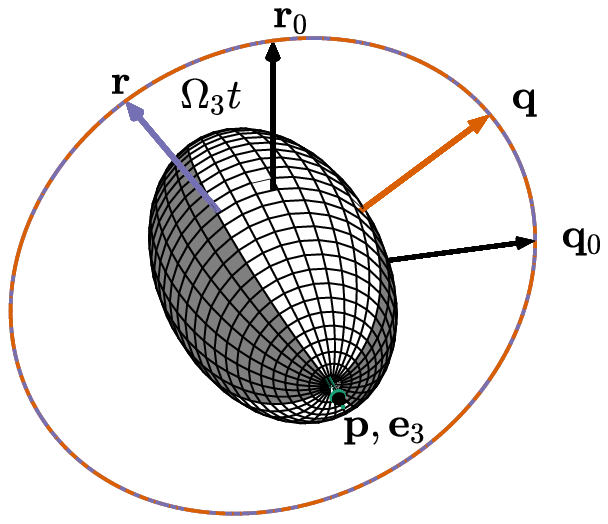}
	\caption{Spinning}
	\label{fig:motion:spinning}
\end{subfigure}
\begin{subfigure}[b]{0.32\linewidth}
	\includegraphics[clip,width=\linewidth]{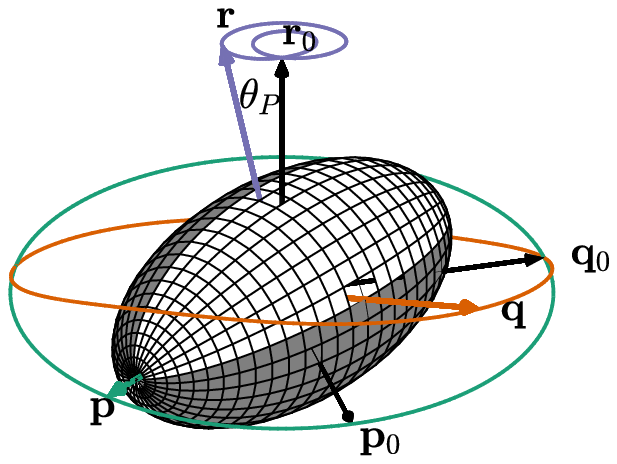}
	\caption{2D tumbling}
	\label{fig:motion:tumbling}
\end{subfigure}
\begin{subfigure}[b]{0.32\linewidth}
	\includegraphics[clip,width=\linewidth]{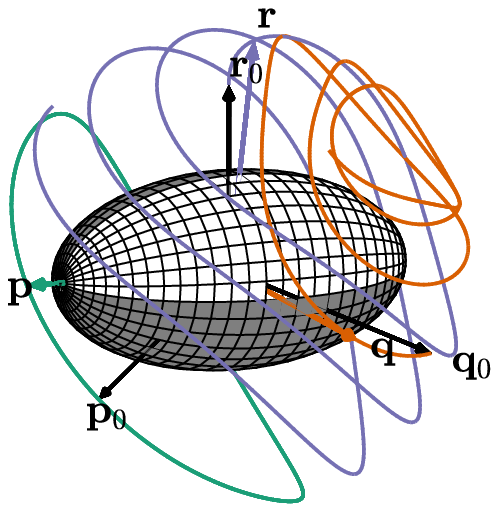}
	\caption{3D tumbling}
	\label{fig:motion:tumbling3d}
\end{subfigure}
\caption{Motion of a freely suspended spheroid: (\subref{fig:motion:spinning}) spinning, (\subref{fig:motion:tumbling}) 2D tumbling and (\subref{fig:motion:tumbling3d}) 3D tumbling. For 2D tumbling and spinning, the motion is periodic, whereas in for 3D tumbling only the motion of the symmetry axis is necessarily periodic.}
\label{fig:motion}
\end{figure}

In case 1a and 2a (spinning), the particle rotates around a fixed axis $\bm{p}=\bm{e}_3$, which is illustrated in figure \ref{fig:motion:spinning}. 
For this spinning motion, the unit vectors of the body frame rotate as
\begin{equation}
\label{eqn:body-frame-steady}
\bm{p} = \bm{e}_3
~,~
\bm{q} = \cos(\Omega_3 t) \bm{q}_0 + \sin(\Omega_3 t) \bm{r}_0
~,~
\bm{r} =-\sin(\Omega_3 t) \bm{q}_0 + \cos(\Omega_3 t) \bm{r}_0
\end{equation}
To obtain the time average velocity gradient, we can substitute (\ref{eqn:body-frame-steady}) into (\ref{eqn:vgt-body-frame}) and take the time average over one revolution, whose period is $T = 2\pi/\Omega_3$.
After a little algebra, we obtain
\begin{equation}
\label{eqn:average-gradient-1a}
\bar{\bm{A}} = \frac{1}{T} \int_{0}^T \tr{\bm{R}}(\bm{G}-[\bm{\Omega}_\infty]_\times)\bm{R} \mathrm{d}t 
= E_3 \begin{bmatrix}
1 & 0 & 0 \\
0 & -\frac{1}{2}  & 0 \\
0 & 0 & -\frac{1}{2} 
\end{bmatrix}
\end{equation}
where
\begin{equation}
\label{eqn:E3}
E_3 = \tr{\bm{e}_3}\bm{E}\bm{e}_3 = \frac{\lambda^2+1}{\lambda^2-1} \sigma
\end{equation}
is the rate of strain perceived by the particle along its fixed axis of rotation.
Thus, the average velocity field experienced by a particle steadily rotating about its axis is an axisymmetric strain, whose magnitude is given by the component of strain along the rotation axis.

In case 1b (resting), the particle axis aligns with $\bm{e}_3$, but the rotation rate about this axis vanishes. 
The average velocity field in the body frame is then simply 
\begin{equation}
\label{eqn:average-gradient-1b}
\bar{\bm{A}} = \bm{RGR}^{T}.
\end{equation}
The average velocity gradient $\bar{\bm{A}}$ can be specified with three parameters, up to an arbitrary scaling and rotation about the symmetry axis of the particle.
To see this, we note that $\bm{G}$ has nine components with eight degrees of freedom, since continuity requires $\trace{\bm{G}} = 0$.
The requirement that $\bm{\Omega} = 0$ imposes the constraint
\begin{equation}
\label{eqn:fixed-point}
\bm{\omega} = -2\frac{\lambda^2-1}{\lambda^2+1} \bm{p}\times \bm{E}\bm{p} 
\end{equation}
reducing the number of unknowns by three.
The choice of scale introduces another redundant parameter; our non-dimensionalisation requires $E_{ij}E_{ij} = 1$.
Finally, we note that rotations about the particle's axis are trivial, since the particle is axisymmetric.

In case 2b (2D tumbling), the motion of the particle is more complex, but remains periodic, as illustrated in figure \ref{fig:motion:tumbling}.
From (\ref{eqn:p-case2b}) it follows that the symmetry axis $\bm{p}$ rotates in a plane spanned by vectors $\bm{a},\bm{b}$ with period $T = 2\pi/\omega$.
Thus the rotation $\bm{Q}(t) = \bm{Q}_2\bm{Q}_1$ of the body frame $\bm{R}(t) = \bm{Q}\bm{R}_0$ can be composed of as a rotation $\bm{Q}_1(t)$ about the axis $\bm{a}\times\bm{b}$ (mapping $\bm{p}_0$ onto $\bm{p}$), followed by a rotation $\bm{Q}_2(t)$ about the axis $\bm{p} = \bm{Q}_1\bm{p}_0$ by an angle 
\begin{equation}
\label{eqn:theta_p}
\theta_p(t) = \int_{0}^{t} \frac{1}{2} \bm{\omega}\cdot \bm{p}(t') \mathrm{d}t'
\end{equation}
relative to the plane normal.
By inspection of (\ref{eqn:theta_p}) and (\ref{eqn:p-case2b}), we see that $\theta_p(T) = 0$, since $\bm{p}(t) = -\bm{p}(t+T/2)$. 
Thus, $\bm{R}(t)$ must be periodic.
The average perceived velocity gradient can then be evaluated analogously to (\ref{eqn:average-gradient-1a}), but the resultant expression is much more complicated.
It suffices to remark that the average strain $\bar{E}$ and vorticity $\bar{\omega}$ components of the average \emph{perceived} velocity gradient $\bar{\bm{A}}$ must also satisfy (\ref{eqn:fixed-point}), since in the body frame, the apparent rotation rate of the body must also be zero.
It follows that in case 2b, as in case 1b, the average perceived velocity gradient may also be described by only three parameters, up to a trivial rotation and choice of scale.

In case 3 (3D tumbling), the particle motion becomes fully three dimensional, as illustrated in figure \ref{fig:motion:tumbling3d}.
Only the motion of its symmetry axis is necessarily periodic; the equatorial axes $\bm{q}$ and $\bm{r}$ may precess around and do not necessarily trace a path with the same period.
Thus, we cannot expect the flow in the particle frame to be time periodic, since $\bm{q}$ and $\bm{r}$ point in a new direction at the start of each cycle, and we cannot expect to apply (\ref{eqn:sherwood-solution}) in this special case.

We recall that the derivation of (\ref{eqn:sherwood-solution}) required pathlines of the flow to be open.
Far from the particle, pathlines follow $\dot{\bm{y}} = \bm{Gy}$, with solution $\bm{y} = \exp(\bm{G}t)\bm{y}_0$.
By a similar argument to that above, we see that when $\bm{G}$ has eigenvalues $0,\pm \mathrm{i}\omega$, the pathlines are closed.
This provides a necessary condition to apply (\ref{eqn:sherwood-solution}).
The 3D tumbling orbits identified by \citet{Jeffery1922} happen to correspond to this case, which further precludes application of (\ref{eqn:sherwood-solution}) to the case of Jeffery orbits.

To summarise: by an analysis of the stable attractors of (\ref{eqn:pdot}) we can construct five cases of the particle motion categorised by the eigenvalues of $\bm{K}$ (\ref{eqn:p}) as shown in table \ref{tbl:spheroid-motion}. 
In cases 1a, 1b and 2a, the particle orients itself along a fixed axis corresponding to the eigenvector $\bm{K}$ with largest real eigenvalue.
In cases 2b and 3, the particle undergoes a limit cycle oscillation.
In cases 1a, 1b, 2a and 2b, the relative flow field is steady or periodic.
In cases 1a, 2a (spinning) the average perceived velocity gradient over one period is an axisymmetric strain, because the particle rotates steadily about its axis.
In cases 1b and 2b (resting or 2D tumbling), the average perceived velocity gradient satisfies (\ref{eqn:fixed-point}), which can be specified in terms of three parameters up to a trivial rotation and choice of scale.

\subsection{The fluid motion near the particle}
\label{sec:stokes-flow}

In the body frame, the surface at $\xi = 0$ is stationary, so to leading order in $\xi$ the velocity is
\begin{equation}
\bm{u} = \xi \frac{\partial \bm{u}}{\partial \xi}\Big|_{\xi = 0} + O(\xi^2)
\end{equation}
After differentiation of Jeffery's solution for the relative velocity field for an arbitrary ellipsoid \citep{Jeffery1922}, rewritten in the more compact matrix notation used by \citet{Kim1991}, we find that the velocity gradient normal to the surface is given by
\begin{equation}
\label{eqn:dudn}
\frac{\partial \bm{u}}{\partial \xi}\Big|_{\xi=0}
= \bm{\Phi}\bm{h}_\xi - \bm{h}_\xi\left(\bm{h}_\xi \cdot (\bm{\Phi}\bm{h}_\xi\right))
= \bm{w}
\end{equation}
where $\bm{h}_\xi = \bm{Dx}/|\bm{Dx}|$ is the unit vector surface normal, $\bm{D}$ is a diagonal matrix whose entries are $a_i^{-2}$ and
\begin{equation}
\label{eqn:Phi}
\bm{\Phi} = 
\frac{3}{4\pi a_1 a_2 a_3}\bm{S}.
\end{equation}
The expression for the stresslet $\bm{S}$ is more involved.
We remark here that it is a symmetric matrix with $\trace{\bm{S}} = 0$ whose elements are a linear combination of the components of the velocity gradient $\bm{A}$ with coefficients determined by geometry-dependent ``resistance functions''.
Complete expressions for $\bm{S}$ can be found in \citet{Kim1991}.
Equation (\ref{eqn:dudn}) is valid for the general case of torque-free, tri-axial ellipsoids.
The action of a net torque upon the body adds an additional skew-symmetric component to (\ref{eqn:Phi}) which is not treated here.

Equation (\ref{eqn:dudn}) now defines the surface streamlines, which are everywhere tangent to the surface velocity gradient $\bm{w}$.
The surface streamlines and their critical points are illustrated for an arbitrary linear shear in figure \ref{fig:coordinates}.
From (\ref{eqn:dudn}) we see that critical points occur where the surface normal $\bm{h}_\xi$ coincides with an eigenvector $\pm\bm{q}_i$ of the surface gradient tensor $\bm{\Phi}$.
Since $\bm{\Phi}$ is a symmetric matrix, its eigenvalues are all real and its eigenvectors orthogonal.
In general, $\bm{\Phi}$ has three distinct eigenvalues ordered $\phi_1 < \phi_2 < \phi_3$ and there are six critical points.
In special cases, $\bm{\Phi}$ has a repeated eigenvalue and there is a locus of critical points.

The precise nature of the critical points can be seen by introducing the surface potential
\begin{equation}
\label{eqn:surface-potential}
\phi(\bm{x}) = \bm{h}_\xi \cdot (\bm{\Phi}\bm{h}_\xi) = \frac{\tr{\bm{x}}\bm{D\Phi Dx}}{\tr{\bm{x}}\bm{D}^2\bm{x}}
\end{equation}
which has the property that 
\begin{equation}
\nabla \phi = \frac{2\bm{Dw}}{\tr{\bm{x}}\bm{D}^2\bm{x}}.
\end{equation}
The gradient of $\phi$ measured along surface streamlines is $\bm{w}\cdot\nabla \phi > 0$ everywhere on the surface except at the critical points $\bm{w}=0$ where the surface streamlines terminate.
In other words, the potential $\phi$ always increases as one moves along a surface streamline.
The potential takes a minimum value of $\phi = \phi_1$ when $\bm{h}_\xi = \pm \bm{q}_1$ (red marker in figure \ref{fig:coordinates}), i.e. surface streamlines originate from a ``source'' $\phi = \phi_1$.
It takes a maximum value of $\phi = \phi_3$ when $\bm{h}_\xi = \pm \bm{q}_3$ (blue marker), i.e. surface streamlines terminate at a ``sink'' $\phi = \phi_3$.
Lastly, when the eigenvalues are distinct, all streamlines must pass through the contour $\phi = \phi_2$, except at points $\bm{h}_\xi = \pm \bm{q}_2$, which are saddle points (green marker).
When there is a repeated eigenvalue, either $\phi_2 = \phi_1$ or $\phi_2 = \phi_3$, so there exists a locus of points where streamlines originate (or terminate).
It should be noted that the definition of $\phi$ (\ref{eqn:surface-potential}) and its properties extend also to torque-free, tri-axial ellipsoids.

\subsection{Surface flux under spinning motion}
\label{sec:axisymmetric-flux}

We now proceed to evaluate the surface streamlines about a spheroid aligned with the axis of an axisymmetric straining flow.
This configuration corresponds to the mean flow field about a spinning spheroid.
In the body frame, the velocity gradient tensor is a diagonal matrix whose elements are $A_{11}/2 = -A_{22} = -A_{33} = E_3$.
Likewise, the surface velocity gradient tensor $\bm{\Phi}$ is also diagonal matrix with $\Phi_{11}/2 = -\Phi_{22} = -\Phi_{33} = \beta E_3/2$, where the geometry-dependent prefactor $\beta$ is given by
\begin{equation}
\frac{1}{\beta} = 
\frac{3}{4} \int_{0}^{\infty} \frac{ac^2 t}{(a^2+t)^{3/2}(c^2+t)^2} \mathrm{d}t.
\end{equation}

For this configuration, the surface streamlines are axisymmetric, originating at the ``pole'' and terminating at the ``equator''. 
Therefore, we can use an ellipsoidal polar coordinate system to describe the surface, identifying $\eta$ with the polar angle between the symmetry axis and a point on the surface and $\zeta$ with the azimuthal angle.
We parametrise a point near the surface as
\begin{equation}
\bm{x} = 
\begin{bmatrix}
a\cos\eta \\
c\sin\eta\cos\zeta \\
c\sin\eta\sin\zeta
\end{bmatrix}
+\xi\bm{h}_\xi.
\end{equation}
so that the covariant coordinate vectors at the surface are
\begin{equation}
\bm{h}_\xi 
= \frac{1}{h_\xi} \begin{bmatrix}
c\cos\eta \\
a\sin\eta \cos\zeta \\
a\sin\eta \sin\zeta \\
\end{bmatrix} \\
,~~
\bm{h}_\eta 
= \begin{bmatrix}
-a \sin\eta \\
c\cos\eta \cos\zeta \\
c\cos\eta \sin\zeta \\
\end{bmatrix} 
,~~
\bm{h}_\zeta 
= \begin{bmatrix}
0 \\
-c\sin\eta \sin\zeta \\
+c\sin\eta \cos\zeta \\
\end{bmatrix} 
\end{equation}
with $h_\xi$ chosen to ensure that $\bm{h}_\xi$ is a unit vector.
Then it follows that
\begin{equation}
\label{eqn:u-axisymmetric}
\begin{aligned}
u_\zeta = 0
,~~
u_\eta = \frac{3}{2} \xi |\Phi_{11}| 
\frac{ac \cos\eta\sin\eta}{(a^2\sin^2\eta + c^2\cos^2\eta)^{3/2}}
= F\xi
\end{aligned}
\end{equation}
as required and
\begin{equation}
\label{eqn:rho-axisymmetric}
\rho = c \sin\eta(a^2 \sin^2\eta + c^2\cos^2\eta)^{1/2}.
\end{equation}
Substituting (\ref{eqn:u-axisymmetric}) and (\ref{eqn:rho-axisymmetric}) into (\ref{eqn:sherwood-solution}), we obtain
\begin{equation}
\begin{aligned}
\int_{\eta_0}^{\eta_1} 
{\rho}^{3/2}{F}^{1/2} \mathrm{d}\eta
&= \int_{0}^{\pi/2} \left(
\frac{3}{2}\beta |{E}_{3}| {a}{c}^4 \cos\eta \sin^4\eta
\right)^{1/2}\mathrm{d}\eta \\
&= \left(\frac{\pi {a}{c}^4 \beta |{E}_{3}|}{6}\right)^{1/2} \frac{\Gamma(\frac{7}{4})}{\Gamma(\frac{9}{4})}
\end{aligned}
\end{equation}
and thus
\begin{equation}
\label{eqn:sherwood-axisymmetric}
\begin{aligned}
\Sh &= 0.566 ({a}{c}^4 \beta)^{1/3} |E_3|^{1/3} \Pe^{1/3} \\
    &= \alpha_{||}\left(\frac{|E_3|^* r}{\kappa}\right)^{1/3}
\end{aligned}
\end{equation}
where $\alpha_{||}(\lambda) = 0.566 ({a}{c}^4 \beta)^{1/3}$ represents the dependence of the surface flux upon geometry and $|E_3|^*r/\kappa$ is the P{\'e}clet number based on the strain perceived along the axis of rotation.

The geometry dependent prefactor $\alpha_{||}$ has a relatively weak dependence upon the aspect ratio of the spheroid, varying between $0.762$ to $1.042$ over the interval $1/20 \le \lambda \le 20$.
In contrast, from the definition of the characteristic shear rate $E^*$, $|E_3|$ may vary over the interval $0 < |E_3| \le 2/\sqrt{6}$, depending on the particular configuration of the background strain and the axis of the particle.
Therefore, the alignment of the particle with the background velocity gradient plays a significant role in determining the surface flux for spinning particles.
This is the phenomenon of the partial suppression of convection by rotation first identified by \citet{Batchelor1979}.

Some caution must be taken in utilising the result of (\ref{eqn:sherwood-axisymmetric}).
In our derivation of (\ref{eqn:sherwood-solution}), we have assumed that the boundary layer thickness remains small in comparison to radius of curvature of the surface, so that cross-surface diffusion and higher-order convection terms are negligible.
Yet, when the body is made infinitely slender, this assumption is violated.
We expect that higher order corrections to (\ref{eqn:sherwood-solution}) are required for slender bodies at moderate P{\'e}clet number.

\subsection{Surface flux under tumbling and resting motion}
\label{sec:general-flux}

Under tumbling and resting motion, convenient expressions for the surface streamlines are no longer easy to find by hand.
To proceed in this general case, we adopt a numerical approach to parametrise the surface in terms of coordinates $\bm{x} = \bm{x}(0, \eta, \zeta)$.
Essentially, the task is to draw a set of $j = 1 \ldots n_\zeta$ surface streamlines covering the body, label each streamline with a unique $\zeta = \zeta_j$ and evaluate the position at $i = 1 \ldots n_\eta$ different locations $\eta = \eta_i$ along the streamline.
Then we can create a $n_\eta \times n_\zeta$ mesh of points $\bm{x}_{i,j} = \bm{x}(0,\eta_i,\zeta_j)$ upon the surface to numerically approximate the metric $\rho$, which can be used to numerically integrate (\ref{eqn:sherwood-solution}).

We shall now construct a suitable definition of the streamwise coordinate $\eta$.
We require the surface streamlines be tangent to the surface velocity gradient $\bm{w}$ (\ref{eqn:dudn}), so $\bm{h}_\eta$ must be of the form
\begin{equation}
\label{eqn:h-eta}
\bm{h}_\eta 
= \frac{\mathrm{d}\bm{x}}{\mathrm{d}\eta} 
= \frac{\bm{w}}{\bm{w}\cdot \nabla \eta}
\end{equation}
which has $\nabla \eta \cdot \bm{h}_\eta = 1$ and $\bm{h}_\eta \cdot \bm{h}_\xi = 0$, as required.
Furthermore, we require that $\eta$ should increase monotonically along surface streamlines.
We have seen in \S\ref{sec:stokes-flow} that the surface potential $\phi$ has this property.
Therefore, we identify $\phi$ (\ref{eqn:surface-potential}) with the streamwise coordinate $\eta$.

We require a definition of the $\zeta$ coordinate.
Since $\zeta$ is constant along surface streamlines and every surface streamline passes through $\eta = \phi_2$, $\zeta$ can be thought of as a coordinate along the curve $\bm{x}_0(\zeta) = \bm{x}(0, \phi_2, \zeta)$.
Along $\eta = \phi_2$, from (\ref{eqn:surface-potential}) we derive that the unit surface normal satisfies
\begin{equation}
\bm{h}_\xi\cdot \bm{q}_1 
\pm \sqrt{\frac{\phi_2 - \phi_3}{\phi_2 - \phi_1}} (\bm{h}_\xi\cdot \bm{q}_3) = 0
\end{equation}
so the constraint
\begin{equation}
\label{eqn:zeta}
\zeta = \bm{h}_\xi \cdot \bm{q}_2
\end{equation}
describes a point along $\bm{x}_0(\zeta)$.
The surface can split into four quadrants, depending upon the basin of attraction of the surface streamlines.
To wit, the behaviour in the limits $\bm{h}_\xi \rightarrow \pm \bm{q}_1$ as $\eta \rightarrow \phi_1$ and $\bm{h}_\xi \rightarrow \pm\bm{q}_3$ as $\eta \rightarrow \phi_3$ forms our classification.
Therefore, in each quadrant, the coordinates $(\eta,\zeta) \in [-\phi_1, \phi_3] \times [-1, 1]$ uniquely define a point on the surface.
This definition is general for torque-free, tri-axial ellipsoids.

We now outline the procedure to evaluate (\ref{eqn:sherwood-solution}) numerically.
For each surface quadrant, we choose a set of points $\bm{x}_{0,j} = \bm{x}(0, \phi_2, \zeta_j)$ which lie on the ellipsoid surface $\xi = 0$ and satisfy $\eta(\bm{x}) = \phi_2$ (\ref{eqn:surface-potential}).
We numerically integrate (\ref{eqn:h-eta}) from the initial condition $\bm{x}=\bm{x}_{0,i}$ at $\eta = \phi_2$ to $\eta = \eta_j$, which yields a mesh of points $\bm{x}_{i,j} = \bm{x}(0,\eta_i,\zeta_j)$ upon the particle surface. 
Measuring the surface area $\delta S_{ij}$ of the region enclosed over $[\eta_i,\eta_{i+1}] \times [\zeta_i,\zeta_{i+1}]$, we approximate the surface area density for this surface element as
\begin{equation}
\rho_{ij} \approx \delta S_{ij} (\eta_{i+1}-\eta_i)(\zeta_{i+1}-\zeta_i)
\end{equation}
which is, roughly speaking, a ``cell average'' of $\rho$.
The velocity component $F$ (\ref{eqn:F}) is evaluated at $\bm{x}_{i,j}$ from (\ref{eqn:surface-potential}) and (\ref{eqn:dudn}).
Then, the integral (\ref{eqn:sherwood-solution}) is evaluated as a Riemann sum over the surface elements.

A technical point remains that $\eta_i$ and $\zeta_j$ should be chosen so that the surface is densely covered in mesh points $\bm{x}_{i,j}$.
To achieve this, we can generate a uniform sampling of seed points on the surface.
These seed points can be integrated (\ref{eqn:h-eta}) numerically towards the curve $\bm{x}_0$ to construct $\zeta_j$. 
This guarantees a satisfactory coverage of the surface by the streamlines. 
The $\eta_i$ can be chosen as
\begin{equation}
\eta_i =
\begin{cases}
\phi_1 \cos^2\nu_i + \phi_2 \sin^2\nu_i & \text{ if } \nu_i < 0 \\ 
\phi_2 \cos^2\nu_i + \phi_3 \sin^2\nu_i & \text{ if } \nu_i > 0 
\end{cases}
\end{equation}
for $\nu_i$ evenly distributed on the interval $[-\pi/2,\pi/2]$.

\subsection{Surface flux in rotation dominated flow}
\label{sec:largew}

It is useful to consider the special case $|\bm{\omega}|\rightarrow \infty$ where the vorticity is large and vortex stretching is non-zero ($\tr{\bm{\omega}}\bm{E\omega} \ne 0$).
In this case, it can be shown that the eigenvalues of $\bm{K}$ are complex, its eigenvectors lie parallel and perpendicular to the direction of the vorticity $\bh{\omega} = \bm{\omega}/|\bm{\omega}|$ and the particle rotates with constant angular velocity $\bm{\Omega} \rightarrow \bm{\omega}/2$.
Therefore, the motion either corresponds to case 2a (spinning) or case 2b (tumbling).
In the spinning case, the particle rotates with its symmetry axis parallel to the vorticity vector $\bm{p} = \bh{\omega}$.
In the tumbling case, the particle rotates with its symmetry axis always orthogonal to the vorticity vector, e.g. $\bm{r} = \bh{\omega}$.
The motion of the body frame is therefore analogous to (\ref{eqn:body-frame-steady}).

The configuration can be inferred from the exact eigenvalue relationship
\begin{equation}
\lambda_1 \lambda_2 \lambda_3 
= -2\sigma(\sigma^2 + \omega^2)
= \frac{\gamma^3}{3}\trace{\bm{E}^3} 
+ \frac{\gamma}{4} \tr{\bm{\omega}}\bm{E\omega}
\end{equation}
where $\gamma = (\lambda^2-1)/(\lambda^2+1)$ is the shape co-factor.
In the limit $|\bm{\omega}|\rightarrow \infty$, $\sigma \rightarrow -\gamma E_\omega/2$, where $E_{\omega} = \tr{\bh{\omega}} \bm{E} \bh{\omega}$ is the strain rate perceived in the direction of vorticity.
Therefore, following \S\ref{sec:particle-motion}, we identify that the particle is aligned in the parallel configuration when $\gamma E_\omega > 0$ and the orthogonal configuration when $\gamma E_\omega < 0$.
It follows that, as in (\ref{eqn:average-gradient-1a}), the average perceived velocity gradient is
\begin{equation}
\overline{\bm{A}}
=
E_{\omega}
\begin{bmatrix}
1 & 0 & 0 \\
0 & -\frac{1}{2} & 0 \\
0 & 0 & -\frac{1}{2}
\end{bmatrix}
~\textrm{when}~~
\gamma E_\omega > 0
~~\textrm{or}~~
E_{\omega}
\begin{bmatrix}
-\frac{1}{2} & 0 & 0 \\
0 & -\frac{1}{2} & 0 \\
0 & 0 & 1
\end{bmatrix}
~\textrm{when}~~
\gamma E_\omega < 0.
\end{equation}

In both cases, the average perceived velocity gradient is an axisymmetric strain.
However, the alignment of the symmetry axis of the spheroid with this strain depends upon the sign of the shape co-factor and vortex stretching. 
As a result, we evaluate the surface flux as
\begin{equation}
\label{eqn:Sh-largew}
\Sh = \left\{\begin{matrix}
\alpha_{||} \Pe_{\omega}^{1/3} & \gamma E_\omega > 0 \\ 
\alpha_{\perp} \Pe_{\omega}^{1/3} & \gamma E_\omega < 0 \\
\end{matrix}\right.
\end{equation}
where $\Pe_{\omega} = E_\omega^* r/ \kappa$ is the P{\'e}clet number based on the vortex stretching, $\alpha_{||}(\lambda)$ is given by (\ref{eqn:sherwood-axisymmetric}) and $\alpha_\perp(\lambda)$ is obtained using the numerical procedure outlined in \S\ref{sec:general-flux}.

\begin{figure}%
\centering
\includegraphics[width=0.5\textwidth,clip]{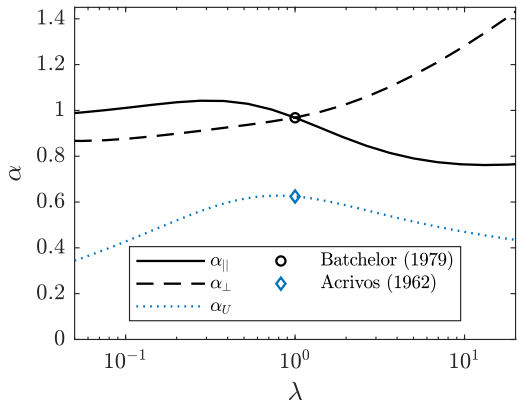}
\caption{Variation in the mass transfer coefficient of a spheroid of varying aspect ratio but constant surface area in axisymmetric straining flow (black lines) and uniform flow (blue line). The black circle shows Batchelor's result for a sphere in axisymmetric strain \citep{Batchelor1979}, whilst the black diamond shows the equivalent result for a sphere in uniform flow \citep{AcrivosTaylor1962}.}
\label{fig:alpha-vs-lambda}
\end{figure}

We have plotted the geometry dependent prefactors $\alpha_{||}$ and $\alpha_{\perp}$ in figure \ref{fig:alpha-vs-lambda}.
The marker shows Batchelor's result $\Sh = 0.968 \Pe_\omega^{1/3}$ for the case of a sphere in rotation dominated flow \citep{Batchelor1979}.
We observe that, for the parallel-aligned configuration $\alpha_{||}$, the variation in the prefactor with $\lambda$ is relatively modest, corresponding to a variation in the surface flux of between $-21.4\%$ and $+7.7\%$ relative to that of a sphere with equivalent surface area.
In contrast, there is a stronger variation observed for the orthogonal configuration $\alpha_{\perp}$, which exhibits an equivalent variation of $-10.5\%$ to $+53\%$ over the range shown. 
From (\ref{eqn:Sh-largew}), we see that the surface flux is proportional to the lower branches of the two black curves in vortex stretching $E_\omega > 0$ and the upper branches in vortex compression $E_\omega < 0$.
It follows that the \emph{sign} of the vortex stretching can influence the surface flux.
In particular, the surface flux to prolate spheroids is significantly increased under vortex compression.

The shape dependence for spheroids in axisymmetric strain may be contrasted to the shape dependence observed for axisymmetric, uniform flow around a fixed spheroid shown in figure \ref{fig:alpha-vs-lambda} \citep{AcrivosTaylor1962,Sehlin1969,Dehdashti2020}.
Here, the surface flux exhibits the same scaling $\Sh = \alpha_U \Pe_U^{1/3} + O(1)$, where $\Pe_U=U_\infty^* r/\kappa$ and $U_\infty^*$ is the magnitude of the relative velocity (slip velocity) between the free-stream and particle.
Whilst it is of limited value to compare the absolute values of $\alpha$, some insight can be obtained by comparing the relative variation of each function.
We observe that for a spherical particle with fixed translation velocity and surface area, flattening or elongating the particle results in a reduction in the surface flux.
Likewise, for a spherical particle in rotation-dominated vortex stretching, flattening or elongating the particle also reduces the surface flux.
However, in rotation-dominated vortex compression, flattening or elongating increases the surface flux.
This observation is relevant to chain formation by marine diatoms, where each cell in a chain must experience an increased nutrient flux per unit area to benefit despite increased competition for nutrients by its neighbours \citep{Pahlow1997}.

One may also observe that rotation dominated straining flow becomes considerably more effective than uniform flow at transferring solute from the particle surface at large aspect ratios.
Of course, this result must depend upon the relative orientation of the free-stream and particle axis, which is not varied here.
Nonetheless, similar comparisons have found utility in determining when sinking or swimming may become a useful strategy for phytoplankton to increase their nutrient flux in turbulent water \citep{KarpBoss1996}.

\subsection{Further extensions} 
\label{sec:generalisations}

Two further generalisations of the surface coordinate system are to cases with non-zero body forces and torques.
This would be necessary, for instance, to capture gyrotactic effects in the nutrient uptake of phytoplankton \citep{Guasto2012} or the behaviour of inertial particles.
Such an extension can be accommodated by a suitable modification of the surface coordinate system, e.g. the inclusion of a body torque adds a skew-symmetric component to $\bm{\Phi}$ in (\ref{eqn:dudn}).
However, in this case, one would also expect the orientation dynamics (and therefore the mean flow around the particle) to change, which would also affect the mean surface flux via the convection-suppression effect.

\section{Comparisons to numerical simulation}
\label{sec:sim}

As a test of the results in in \S\ref{sec:general-theory} and \S\ref{sec:ellipsoid-theory}, we conducted numerical simulations of the unsteady scalar transport around spheroids freely suspended in a linear shear.
We describe our numerical methods in \S\ref{sec:sim:methods} then examine two classes of arbitrary strain and rotational flows in \S\ref{sec:sim:results}.

\subsection{Methods}
\label{sec:sim:methods}

The unsteady convection diffusion equation (\ref{eqn:scalar-transport}) is solved using a second-order finite volume method, subject to the Dirichlet boundary condition $\theta = 1$ on $S_p$ and the von-Neumann boundary condition on the outer boundary of the simulated domain.
Equation (\ref{eqn:scalar-transport}) is discretised on a structured grid in prolate (\ref{eqn:prolate}) or oblate (\ref{eqn:oblate}) spheroidal coordinates $(\mu,\theta,\phi)$, depending upon the particle aspect ratio.
The inner boundary $(\mu_0,\theta,\phi)$ corresponds to the surface of the spheroid oriented in the Cartesian $x$ direction. 
The relative velocity field was evaluated in the body frame based on the coordinate of each cell centre using the expressions given by \citet{Kim1991}.
The relative velocity field satisfies the impermeable, no slip boundary condition $\bm{u}=0$ on $S_p$.

\begin{align}
\label{eqn:prolate}
\bm{x}_{ijk} &= \begin{bmatrix}
f\cosh\mu_i \cos\theta_j \\
f\sinh\mu_i \sin\theta_j \cos\phi_k \\
f\sinh\mu_i \sin\theta_j \sin\phi_k
\end{bmatrix}
~,~
\begin{aligned}
\mu_i &\in [\mu_0, \mu_\infty] \\
\theta_j &\in [0, \pi] \\
\phi_k &\in [0, 2\pi) \\
\end{aligned}
~,~
&f = \sqrt{a^2-c^2} \\
\label{eqn:oblate}
\bm{x}_{ijk} &= \begin{bmatrix}
f\sinh\mu_i \sin \theta_j \\
f\cosh\mu_i \cos \theta_j \cos\phi_k \\
f\cosh\mu_i \cos \theta_j \sin\phi_k
\end{bmatrix}
~,~
\begin{aligned}
\mu_i &\in [\mu_0, \mu_\infty] \\
\theta_j &\in [-\pi/2, \pi/2] \\
\phi_k &\in [0, 2\pi) \\
\end{aligned}
~,~
&f = \sqrt{c^2-a^2}
\end{align}

The mesh resolution and spacing was chosen following the studies of \citet{Pahlow1997} and \citet{KarpBoss1996}.
The mesh is discretised into $150\times 64\times 64$ cells in the $(\mu,\theta,\phi)$ directions respectively, with uniform spacing in the $\theta$ and $\phi$ directions.
Due to the nature of the spheroidal coordinate system chosen, the resolution varies across the surface of the particle and the outer boundary is very slightly oblate or prolate.
The dimensions of the spheroid $a_i = (a,c,c)$ are chosen such that the surface area is $4\pi$, equivalent in surface area to a sphere with unit radius. 
The outer boundary is chosen such that its largest dimension is $100$ and is very nearly spherical, having an aspect ratio between $0.999$ and $1.001$.
To adequately resolve the thin concentration boundary layer, which is of thickness $\delta = \Pe^{-1/3}$, we employ a mesh refinement in the $\mu$ direction such that the grid spacing is $\Delta \mu_{i+1} = R \Delta \mu_i$, where $\Delta \mu_{i+1} = \mu_{i+1}-\mu_i$ is the spacing between adjacent cells in the $\mu$ direction.
The initial spacing $\Delta \mu_1$ is chosen such that the thickness of the largest cell near the particle surface is at most $2\times10^{-4} \max(a,c)$ and the mesh refinement factor $R$ is chosen accordingly.
For the most extreme aspect ratio $\lambda = 16$ ($\lambda = 1/16$) at the largest P{\'e}clet number tested, this yields between $27$ (43) and $70$ (87) cells within a distance $\delta$ from the surface.

The solver is based on the \textit{scalarTransportFoam} solver of \textit{OpenFOAM}, which was modified to solve (\ref{eqn:scalar-transport}) for a time varying flow field.
The convective term in (\ref{eqn:scalar-transport}) is discretised using a standard linear upwind Gaussian integration, and the diffusive term is discretised using a similar linear Gaussian scheme with an explicit non-orthogonal correction to maintain second order accuracy.
Time stepping is performed using an implicit Euler scheme and a time step of $\Delta t = 0.02$.
Simulations were allowed sufficient time for the surface flux to reach a steady (or periodic) state.
Where the particle motion is unsteady, the cycle average mass flux was evaluated over the interval $9T \le t < 10T$.
Where the particle motion is steady, the steady state mass flux was evaluated at $t = 100$.

\subsection{Results and discussion}
\label{sec:sim:results}

In this section, we shall compare the results of our numerical simulations against the asymptotic results derived in \S\ref{sec:general-theory}.

\subsubsection{Pure strain}

As our first test, we consider an arbitrary irrotational background velocity gradient $\bm{G}=\bm{E}$.
The particle motion in this case corresponds to case 1b of \S\ref{sec:particle-motion} and the particle aligns itself with the most extensional (or compressive) direction of strain.
Thus, the relative velocity gradient field is of the form
\begin{equation}
\label{eqn:strain}
\bm{A} = \begin{bmatrix}
\sigma_1 & 0 & 0 \\
0 & \sigma_2 & 0 \\
0 & 0 & \sigma_3 
\end{bmatrix}
\end{equation}
where $\sigma_i$ are the eigenvalues of $\bm{E}$.
The topology of the relative flow field can therefore described by a single parameter $-1 \le s \le 1$ \citep{LundRogers1994}
\begin{equation}
s = \frac{-3\sqrt{6} \sigma_1 \sigma_2 \sigma_3}{(\sigma_1^2 + \sigma_2^2 + \sigma_3^2)^{3/2}}.
\end{equation}
When $s = 1$, the background flow is an axisymmetric extensional flow; $s = 0$ corresponds to a two-dimensional strain and $s = -1$ corresponds to an axisymmetric compression.

\begin{figure}
\centering
\begin{subfigure}[b]{0.49\linewidth}
	\includegraphics[clip,width=\textwidth]{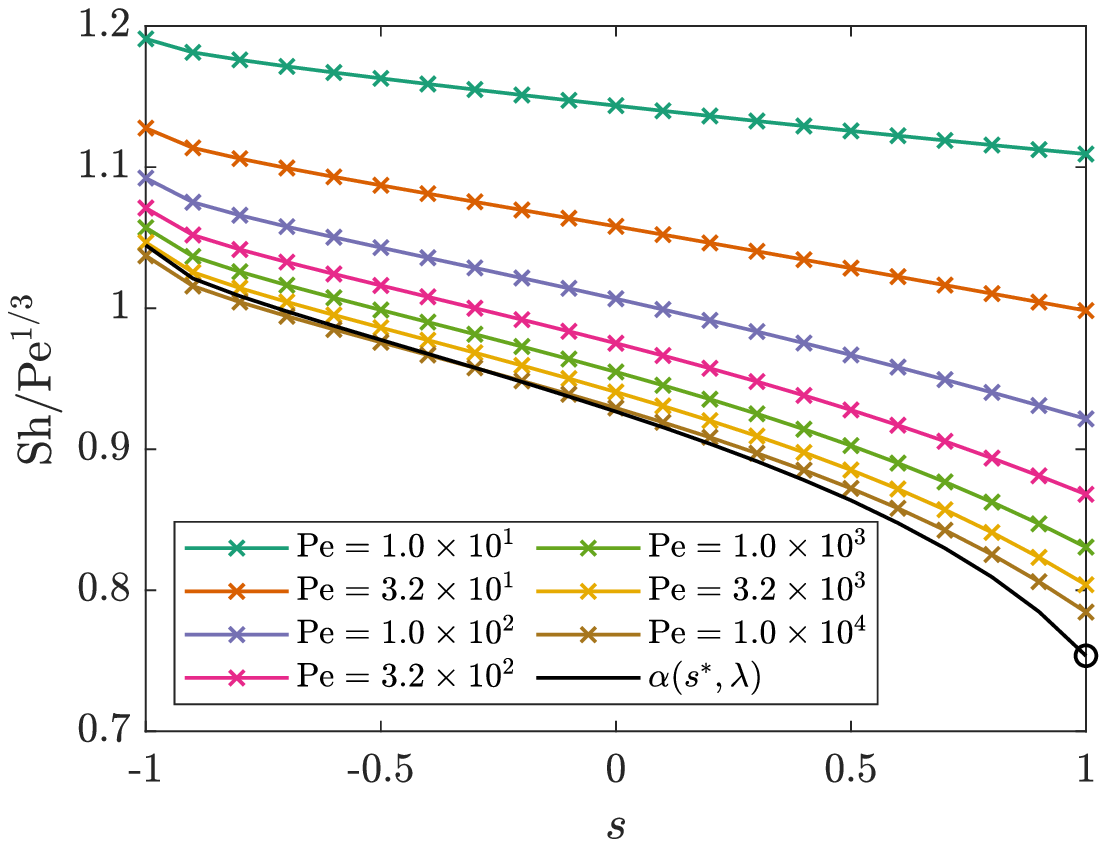}
	\caption{Prolate spheroid, $\lambda=4$}
	\label{fig:strain:a}
\end{subfigure}
\hfill
\begin{subfigure}[b]{0.49\linewidth}
	\includegraphics[clip,width=\textwidth]{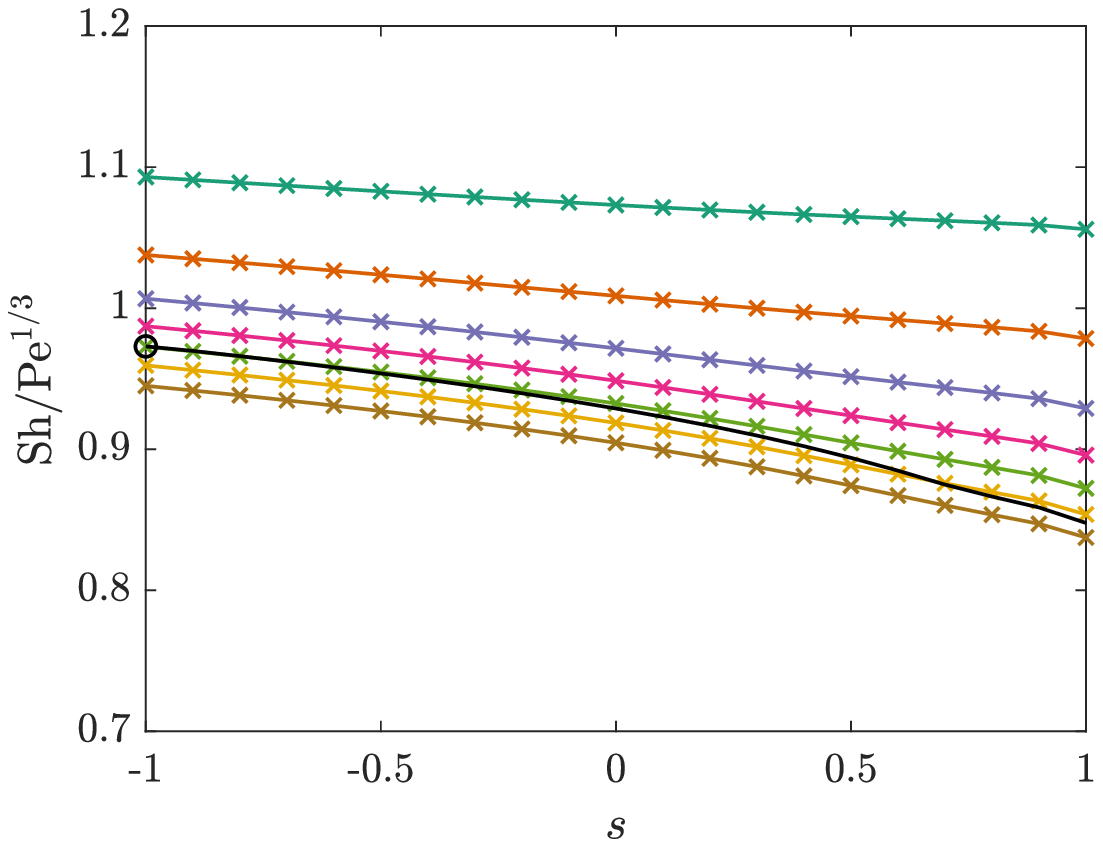}
	\caption{Oblate spheroid, $\lambda=1/4$}
	\label{fig:strain:b}
\end{subfigure}
\caption{Mass transfer coefficient for (\subref{fig:strain:a}) prolate and (\subref{fig:strain:b}) oblate spheroids in a pure straining flow, as a function of the strain topology parameter $s$.
Coloured lines with markers show numerical simulations over $\Pe = 10^1 - 10^4$.
Black lines show the expected scaling coefficient $\alpha(s,\lambda)$ (\ref{eqn:sherwood-solution}).
Black circular markers show the axisymmetric result (\ref{eqn:sherwood-axisymmetric}).}
\label{fig:strain}
\end{figure}

We computed the surface flux from prolate and oblate spheroids in an arbitrary straining flow over a range of $\Pe$.
The result is shown in figure \ref{fig:strain} for cases $\lambda = 4$ and $1/4$.
For both prolate and oblate spheroids, there is a modest variation ($15 - 30\%$) in the mass transfer coefficient as the strain topology is varied, which becomes more pronounced as the P{\'e}clet number is increased.
In contrast, the mass transfer coefficient for a spherical body in a pure straining flow varies by less than $1\%$ over the same range of $s$ \citep{Batchelor1979}.
When the P{\'e}clet number is large, the mass transfer rate approaches the limiting scaling $\Sh = \alpha\Pe^{1/3}$.
The numerical coefficient $\alpha$ is well predicted by (\ref{eqn:sherwood-solution}) and shows the correct qualitative dependence upon $s$. 
At $\Pe = 10^4$, the discrepancy in the predicted mass transfer rate between theory and numerics is within $2.5\%$ for the prolate case and within $3.1\%$ for the oblate case.
This is partly due to the mesh refinement; additional tests at a higher resolution of $300 \times 128 \times 128$ suggest the numerics slightly under-resolve the surface flux by around 2\% for the oblate case, which would improve the agreement seen here.
Nonetheless, the leading order correction term to (\ref{eqn:sherwood-solution}) is $O(1)$, which corresponds to a discrepancy in $\Sh/\Pe^{1/3}$ of $\sim 4.6\%$ at $\Pe = 10^4$ and is comparable to the observed discrepancy.

\begin{figure}
\centering
\includegraphics[clip,width=0.5\textwidth]{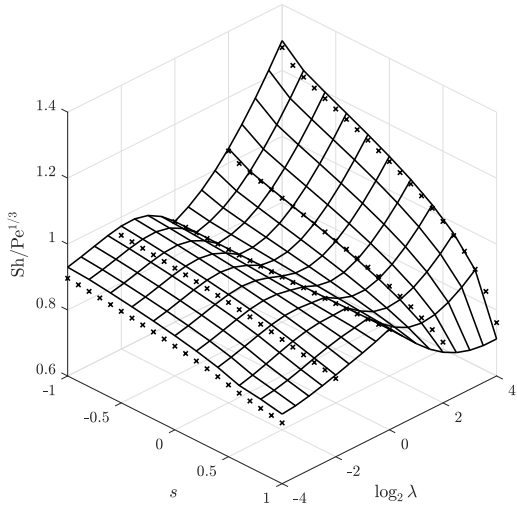}
\caption{Mass transfer coefficient for spheroids in a pure straining flow, as a function of the strain topology parameter $s$ and aspect ratio $\lambda$.
The ruled surface shows the expected scaling coefficient $\alpha(s,\lambda)$ (\ref{eqn:sherwood-solution}).
Black markers show the result of numerical simulations conducted at $\Pe=10^4$.}
\label{fig:strain-surface}
\end{figure}

To examine the role of particle shape, we plot the scaling coefficient $\alpha(s,\lambda)$ over a range of $s$ and $\lambda$ in figure \ref{fig:strain-surface}.
Markers show the value of $\Sh/\Pe^{1/3}$ from our numerical simulations at $\Pe=10^4$ whilst the ruled surface shows the result of (\ref{eqn:sherwood-solution}).
We observe that prolate spheroids tend to experience a larger surface flux than oblate spheroids of equivalent surface area in the same flow.
However, the trend not always clear-cut and is reversed for strain topologies near $s=1$, where spherical particles experience a larger surface flux.

In figure \ref{fig:strain} and figure \ref{fig:strain-surface}, we observe that the surface flux is always larger in axisymmetric compression ($s=-1, \sigma_1 > 0, \sigma_2 = \sigma_3 < 0$) than axisymmetric extension ($s=+1, \sigma_1 < 0, \sigma_2 = \sigma_3 > 0$).
This is remarkable, since both flows correspond to an axisymmetric strain; only the direction of the flow is reversed.
At first glance, this appears to violate Brenner's flow reversal theorem \citep{Brenner1967,Masoud2019}, which states that for an isothermal body in steady flow, the surface flux is preserved under flow reversal $\bm{u} \rightarrow -\bm{u}$.
However, we recall that the stable orientation of a free spheroid shifts under flow reversal.
In this example, prolate spheroids align parallel to the Cartesian $1-$direction in axisymmetric compression ($s=-1$) and orthogonal to it in axisymmetric extension ($s=+1$).
In fact, the flow configuration is identical to the parallel and orthogonal flow configurations discussed in \S\ref{sec:largew} and shown in figure \ref{fig:alpha-vs-lambda}.
Thus, the difference in the average transfer rate between $s=\pm 1$ is due to a change in the stable orientation of the spheroid.

\subsubsection{Spinning and tumbling}

As a numerical test of our result for spinning and tumbling spheroids, we consider a spheroid in a background velocity gradient $\bm{G}=\bm{E}+\bm{W}$ of the form
\begin{equation}
\label{eqn:shear}
\bm{E} = \frac{1}{\sqrt{6}}
\begin{bmatrix}
2 & 0 & 0 \\
0 & -1& 0 \\
0 & 0 & -1
\end{bmatrix}
~,~
\bm{W} = \begin{bmatrix}
0 & -\sin\theta_{\omega} & 0 \\
\sin\theta_{\omega} & 0 & -\cos\theta_{\omega} \\
0 & \cos\theta_{\omega} & 0
\end{bmatrix}
\end{equation}
This corresponds to an axisymmetric straining flow with $|\bm{E}|=1$ and vorticity magnitude $|\omega| = 2$.
By varying the angle $\theta_\omega$ between the background vorticity $\bm{\omega}$ and the $x_1$ axis, we survey different limiting behaviours of particle motion identified in \S\ref{sec:particle-motion} and the relative flow field experienced by the particle depends upon is geometry.
For example, a prolate particle will spin when $\theta_\omega = 0$  (case 2a) whereas it will tumble  for $\theta_\omega = \pi/2$ (case 2b). 
The situation is reversed for an oblate particle.

\begin{figure}
\centering
\begin{subfigure}[b]{0.49\linewidth}
	\includegraphics[clip,width=\textwidth]{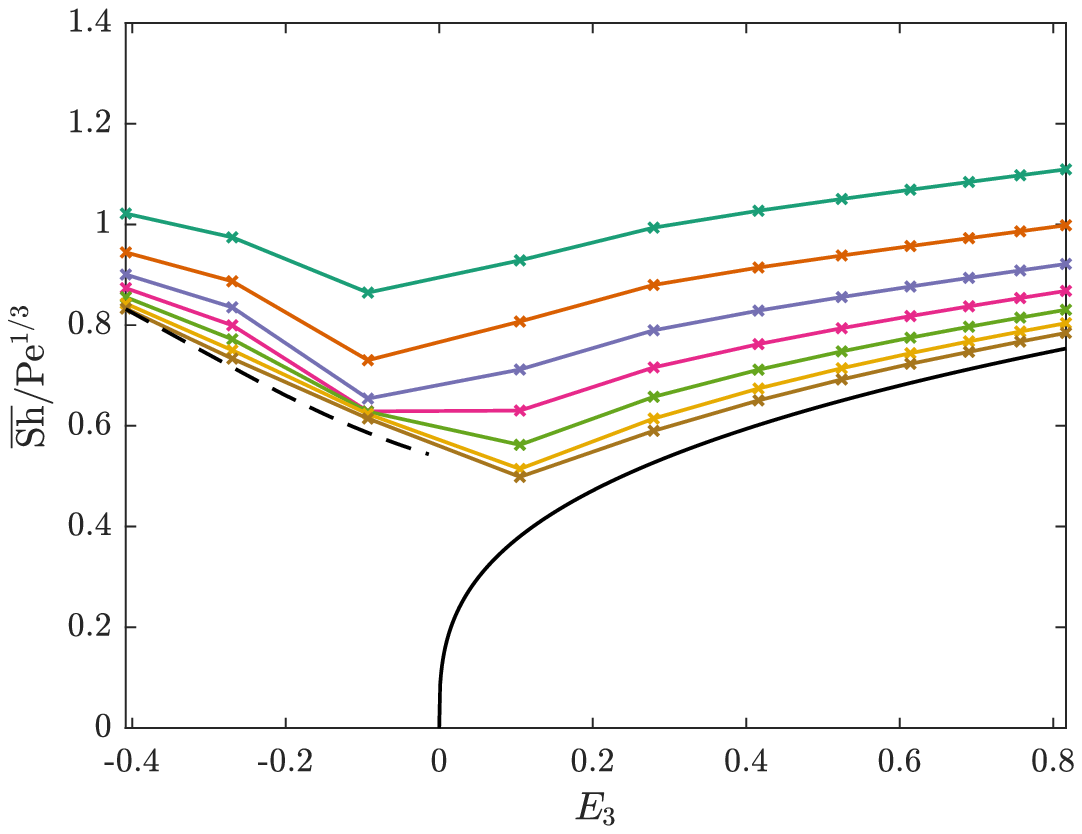}
	\caption{Prolate spheroid, $\lambda=4$}
	\label{fig:spinning-tumbling:a}
\end{subfigure}
\hfill
\begin{subfigure}[b]{0.49\linewidth}
	\includegraphics[clip,width=\textwidth]{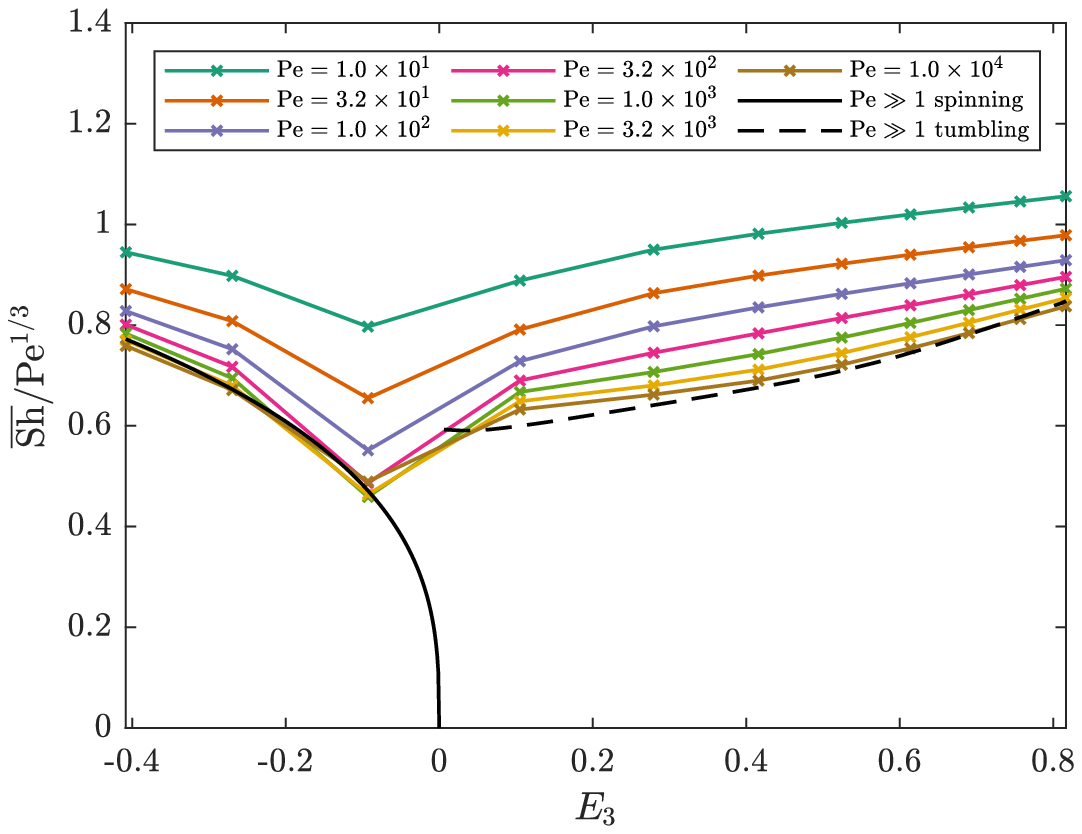}
	\caption{Oblate spheroid, $\lambda=1/4$}
	\label{fig:spinning-tumbling:b}
\end{subfigure}
\caption{Average mass transfer coefficient for (\subref{fig:spinning-tumbling:a}) prolate and (\subref{fig:spinning-tumbling:b}) oblate spheroids in an arbitrary shear (\ref{eqn:shear}), as an implicit function of the mean axial strain rate $E_3$ (\ref{eqn:E3}).
Coloured lines with markers show numerical simulations over $\Pe = 10^1 - 10^4$.
Black solid lines show the expected scaling coefficient for spinning motion (\ref{eqn:sherwood-axisymmetric}).
Black dashed lines show the expected scaling coefficient for tumbling motion (\ref{eqn:sherwood-axisymmetric}).}
\label{fig:spinning-tumbling}
\end{figure}

Figure \ref{fig:spinning-tumbling} shows how the average mass transfer rate varies for prolate ($\lambda = 4$) and oblate $(\lambda = 1/4)$ spheroids as the parameter $\theta_\omega$ is varied.
The dependence is plotted as an implicit function of the axial strain rate $E_3$ (\ref{eqn:E3}) for $\Pe = 10^1 - 10^4$.
We first remark that as the P{\'e}clet number becomes large, the mass transfer rate approaches the scaling predicted by (\ref{eqn:sherwood-axisymmetric}) and (\ref{eqn:sherwood-solution}).
As $\theta_\omega$ is varied, the particle motion switches from spinning to tumbling at $E_3 = 0$ and a pronounced change can be observed in the limiting mass transfer rate.
There is a marked suppression of mass transfer near $E_3 = 0$ as the particle approaches the transition from spinning to tumbling.
This is a demonstration of the suppression of convection by rotation first identified by \citet{Batchelor1979}.
We note, however, that the presence of vorticity does not always suppress the convective transport.
For instance, for the prolate spheroid in figure \ref{fig:spinning-tumbling:a}, tumbling induced by the vorticity component can enhance the convective transfer relative to the equivalent pure strain case (figure  \ref{fig:strain:a}).

Some remarks are in order.
Firstly, the convection suppression/augmentation effect is essentially a hydrodynamic effect: it occurs as the result of the motion of a free spheroid and its alignment with the strain field.
Secondly, this effect can be seen even at moderate P{\'e}clet number and becomes more pronounced as $\Pe$ increases.
This underlines the observation that particle shape influences two factors in determining the mass transfer rate: it determines both the boundary condition for the scalar field and the behaviour of relative flow field.

\section{Conclusion}
\label{sec:conclusions}

In this paper, we have presented a general method to evaluate the average surface flux of solute from a rigid particle of arbitrary shape immersed in an arbitrary, open pathline flow.
The main restrictions upon the shape of the particle are that it contains no sharp edges or regions of extreme curvature, where the thin boundary layer assumption breaks down.
The flow may be steady or time periodic.
When the flow is periodic, the average surface flux is equivalent to that of the same particle embedded in the mean flow field, provided the dimensionless period of the motion is $T \ll \Pe^{1/3}$.
The Sherwood number scales as $\Pe^{1/3}$, with a prefactor $\alpha$ which can be readily obtained through numerical integration once the particle geometry and surface flow field are specified.

We apply this result to compute the surface flux from a small, freely suspended, spheroid in a steady linear shear.
To do so, we compute the relative flow field experienced by the particle, which may be unsteady due to the particle motion.
Through an analysis of Jeffery's equation, we identify four categories of motion: resting, spinning and 2D or 3D tumbling.
The relative flow field is time periodic in the first three cases.
In the spinning case, the average perceived flow field is an axisymmetric strain.
In the 2D tumbling case, the average flow field always corresponds to an equivalent spheroid in steady flow (resting).
We provide a closed form expression (\ref{eqn:sherwood-axisymmetric}) for the surface flux in the spinning case.
We outline the numerical procedure to obtain the surface flux in the resting or 2D tumbling cases.
We also describe a simplification for the case of rotation dominated flow $|\bm{\omega}| \rightarrow \infty$. 
In this limit, there is a larger surface flux under vortex compression when compared to vortex stretching and this increase becomes significant for prolate bodies.
These procedures may serve as the basis for analysing other, more complex geometries, or more complex rigid body dynamics including inertia and gravity.

As a test of these analytical results, we have presented numerical simulations of the scalar transport and surface flux around spheroids in pure straining and a simplified shear flow.
In all cases, we observe good agreement with the expected scaling law and mass transfer coefficient, up to the accuracy of the asymptotic approximation.
In pure straining flows, the surface flux is steady and is prescribed by only three parameters: the P{\'e}clet number, the particle aspect ratio and a parameter describing the strain topology.
In surveying this parameter space, we observe that prolate spheroids tend to experience a greater surface flux than oblate spheroids of equivalent surface area.
When vorticity is present, we observe that the spinning or tumbling of the particle may suppress or augment the convective transfer, due to a realignment of the particle with respect to the average strain field.
Two additional parameters are necessary to characterise the surface flux in all rotational flows and a complete survey of the parameter space is not practical.
However, the space is sufficiently small that it may be readily tabulated for use as a model in a numerical simulation.

We anticipate that our results may find application in modelling the mass transfer from spheroidal or arbitrary particles in numerical simulations of particle laden flows, where models for the interphase mass transfer rate are required to take particle shape into account.
Although the results presented here pertain to steady and time periodic flows, the results may serve in the same spirit in which steady flow mass transfer coefficients are employed to model the transfer rate in unsteady flows \citep{Crowe2012}.
Of particular interest is the mass transfer in turbulent environments, such as turbulent ocean waters home to planktonic osmotrophs \citep{KarpBoss1996} and microplastics \citep{Law2017}.
Here, the adaptation of shape to maximise surface flux may help explain the great diversity in the morphology of osmotrophs \citep{Guasto2012}, or help identify which shapes of microplastics do the most harm. 
Another extension would be to include non-zero body torques or forces.
This would allow, for instance, the consideration of gyrotactic effects in the motion of phytoplankton \citep{Guasto2012}, or inertial effects in the rigid body dynamics of suspended particles.

\section{Acknowledgements}
This project has received funding from the European Union's Horizon 2020 research and innovation programme under the Marie Sk\l{}odowska-Curie grant agreement No 846648.
The authors acknowledge the use of the IRIDIS High Performance Computing Facility, and associated support services at the University of Southampton, in the completion of this work.
The authors report no conflict of interest.

\bibliographystyle{jfm}
\bibliography{rapids}

\begin{thebibliography}{38}
\expandafter\ifx\csname natexlab\endcsname\relax\def\natexlab#1{#1}\fi
\def\au#1{#1} \def\ed#1{#1} \def\yr#1{#1}\def\at#1{#1}\def\jt#1{\textit{#1}}
  \def\bt#1{#1}\def\bvol#1{\textbf{#1}} \def\vol#1{#1} \def\pg#1{#1}
  \def\publ#1{#1}\def\arxiv#1{#1}\def\org#1{#1}\def\st#1{\textit{#1}}

\bibitem[Acrivos(1960)]{Acrivos1960}
{\sc \au{Acrivos, Andreas}} \yr{1960}  \at{{Solution of the laminar boundary
  layer energy equation at high Prandtl numbers}}.  \jt{Physics of Fluids}
  \bvol{3}~(4),  \pg{657--658}.

\bibitem[Acrivos(1980)]{Acrivos1980}
{\sc \au{Acrivos, Andreas}} \yr{1980}  \at{{A note on the rate of heat or mass
  transfer from a small particle freely suspended in a linear shear field}}.
  \jt{Journal of Fluid Mechanics}  \bvol{98}~(2),  \pg{299--304}.

\bibitem[Acrivos \& Taylor(1962)]{AcrivosTaylor1962}
{\sc \au{Acrivos, Andreas} \& \au{Taylor, Thomas~D.}} \yr{1962}  \at{{Heat and
  mass transfer from single spheres in stokes flow}}.  \jt{Physics of Fluids}
  \bvol{5}~(4),  \pg{387--394}.

\bibitem[Batchelor(1979)]{Batchelor1979}
{\sc \au{Batchelor, G.K.}} \yr{1979}  \at{{Mass transfer from a particle
  suspended in fluid with a steady linear ambient velocity distribution}}.
  \jt{Journal of Fluid Mechanics}  \bvol{95},  \pg{369--400}.

\bibitem[Batchelor(1980)]{Batchelor1980}
{\sc \au{Batchelor, G.~K.}} \yr{1980}  \at{{Mass transfer from small particles
  suspended in turbulent fluid}}.  \jt{Journal of Fluid Mechanics}
  \bvol{98}~(3),  \pg{609--623}.

\bibitem[Brenner(1967)]{Brenner1967}
{\sc \au{Brenner, H.}} \yr{1967}  \at{{On the invariance of the heat-transfer
  coefficient to flow reversal in Stokes and potential streaming flows past
  particles of arbitrary shape}}.  \jt{J. Math. Phys. Sci.}  \bvol{1},
  \pg{173--179}.

\bibitem[Bretherton(1962)]{Bretherton1962}
{\sc \au{Bretherton, F.~P.}} \yr{1962}  \at{{The motion of rigid particles in a
  shear flow at low Reynolds number}}.  \jt{Journal of Fluid Mechanics}
  \bvol{14}~(2),  \pg{284--304}.

\bibitem[Clift {\em et~al.\/}(1978)Clift, Grace \& Weber]{Clift1978}
{\sc \au{Clift, M.}, \au{Grace, J.~R.} \& \au{Weber, M.E.}} \yr{1978} {\em
  {Bubbles, Drops and Particles}\/}.  \publ{New York: Academic Press, Inc.}

\bibitem[Crowe {\em et~al.\/}(2012)Crowe, Schwarzkopf, Sommerfeld \&
  Tsuji]{Crowe2012}
{\sc \au{Crowe, C.~T.}, \au{Schwarzkopf, J.~D.}, \au{Sommerfeld, M.} \&
  \au{Tsuji, Y.}} \yr{2012} {\em {Multiphase flows with Droplets and
  Particles}\/}.  \publ{CRC Press}.

\bibitem[Dehdashti \& Masoud(2020)]{Dehdashti2020}
{\sc \au{Dehdashti, Esmaeil} \& \au{Masoud, Hassan}} \yr{2020}  \at{{Forced
  Convection Heat Transfer from a Particle at Small and Large Peclet Numbers}}.
   \jt{Journal of Heat Transfer}  \bvol{142}~(6),  \pg{1--9}.

\bibitem[Feng \& Michaelides(1997)]{Feng1997}
{\sc \au{Feng, Zhi~Gang} \& \au{Michaelides, Efstathios~E.}} \yr{1997}
  \at{{Unsteady Heat and Mass Transfer from a Spheroid}}.  \jt{AIChE Journal}
  \bvol{43}~(3),  \pg{609--614}.

\bibitem[Frankel \& Acrivos(1968)]{FrankelAcrivos1968}
{\sc \au{Frankel, Neil~A.} \& \au{Acrivos, Andreas}} \yr{1968}  \at{{Heat and
  mass transfer from small spheres and cylinders freely suspended in shear
  flow}}.  \jt{Physics of Fluids}  \bvol{11}~(9),  \pg{1913--1918}.

\bibitem[Goddard \& Acrivos(1965)]{GoddardAcrivos1965}
{\sc \au{Goddard, J.D.} \& \au{Acrivos, A.}} \yr{1965}  \at{{Asymptotic
  expansions for laminar forced-convection heat and mass transfer}}.  \jt{J.
  Fluid Mech.}  \bvol{23}~(1965),  \pg{273--291}.

\bibitem[Grinfeld(2013)]{Grinfeld2013}
{\sc \au{Grinfeld, Pavel}} \yr{2013} {\em {Introduction to tensor analysis and
  the calculus of moving surfaces}\/}.  \publ{New York: Springer}.

\bibitem[Guasto {\em et~al.\/}(2012)Guasto, Rusconi \& Stocker]{Guasto2012}
{\sc \au{Guasto, Jeffrey~S.}, \au{Rusconi, Roberto} \& \au{Stocker, Roman}}
  \yr{2012}  \at{{Fluid Mechanics of Planktonic Microorganisms}}.  \jt{Annual
  Review of Fluid Mechanics}  \bvol{44}~(1),  \pg{373--400}.

\bibitem[Gupalo {\em et~al.\/}(1976)Gupalo, Polianin \& Riazantsev]{Gupalo1976}
{\sc \au{Gupalo, Iu.~P.}, \au{Polianin, A.~D.} \& \au{Riazantsev, Iu.~S.}}
  \yr{1976}  \at{{Diffusion to a particle at large Peclet numbers in the case
  of arbitrary axisymmetric flow over a viscous fluid}}.  \jt{Journal of
  Applied Mathematics and Mechanics}  \bvol{40}~(2),  \pg{893--898}.

\bibitem[Gupalo \& Riazantsev(1972)]{Gupalo1972}
{\sc \au{Gupalo, Iu.~P.} \& \au{Riazantsev, Iu.~S.}} \yr{1972}  \at{{Diffusion
  on a particle in the shear flow of a viscous fluid. Approximation of the
  diffusion boundary layer}}.  \jt{Journal of Applied Mathematics and
  Mechanics}  \bvol{36}~(3),  \pg{447--451}.

\bibitem[Jeffery(1922)]{Jeffery1922}
{\sc \au{Jeffery, G.B.}} \yr{1922}  \at{{The Motion o f Ellipsoidal Particles
  Immersed}}.  \jt{Proceedings of the Royal Society A: Mathematical, Physical
  and Engineering Sciences}  \bvol{102}~(715),  \pg{161 -- 179}.

\bibitem[Karp-Boss {\em et~al.\/}(1996)Karp-Boss, Boss \& Jumars]{KarpBoss1996}
{\sc \au{Karp-Boss, L.}, \au{Boss, E.} \& \au{Jumars, P.~A.}} \yr{1996}
  \at{{Nutrient fluxes to planktonic osmotrophs in the presence of fluid
  motion}}.  \jt{Oceanography and Marine Biology: an annual review}  \bvol{34},
   \pg{71--107}.

\bibitem[Ke {\em et~al.\/}(2018)Ke, Shu, Zhang, Yuan \& Yang]{Ke2018}
{\sc \au{Ke, Chunhai}, \au{Shu, Shi}, \au{Zhang, Hao}, \au{Yuan, Haizhuan} \&
  \au{Yang, Dongmin}} \yr{2018}  \at{{On the drag coefficient and averaged
  Nusselt number of an ellipsoidal particle in a fluid}}.  \jt{Powder
  Technology}  \bvol{325},  \pg{134--144}.

\bibitem[Kim \& Karrila(1991)]{Kim1991}
{\sc \au{Kim, Sangtae} \& \au{Karrila, Seppo~J.}} \yr{1991} {\em
  {Microhydrodynamics}\/}.  \publ{Elsevier}.

\bibitem[Kishore \& Gu(2011)]{Kishore2011}
{\sc \au{Kishore, Nanda} \& \au{Gu, Sai}} \yr{2011}  \at{{Momentum and heat
  transfer phenomena of spheroid particles at moderate Reynolds and Prandtl
  numbers}}.  \jt{International Journal of Heat and Mass Transfer}
  \bvol{54}~(11-12),  \pg{2595--2601}.

\bibitem[Law(2017)]{Law2017}
{\sc \au{Law, Kara~Lavender}} \yr{2017}  \at{{Plastics in the Marine
  Environment}}.  \jt{Annual Review of Marine Science}  \bvol{9}~(1),
  \pg{205--229}.

\bibitem[Leal(2012)]{Leal2012}
{\sc \au{Leal, L.~Gary}} \yr{2012}  \at{{Convection Effects in
  Low-Reynolds-Number Flows}}.  \bt{In {\em Advanced Transport Phenomena\/}}, ,
   \vol{vol. 135},  \pg{p. 663}.  \publ{Cambridge University Press}.

\bibitem[Lund \& Rogers(1994)]{LundRogers1994}
{\sc \au{Lund, Thomas~S.} \& \au{Rogers, Michael~M.}} \yr{1994}  \at{{An
  improved measure of strain state probability in turbulent flows}}.
  \jt{Physics of Fluids}  \bvol{6}~(5),  \pg{1838--1847}.

\bibitem[Ma \& Zhao(2020)]{Ma2020}
{\sc \au{Ma, Huaqing} \& \au{Zhao, Yongzhi}} \yr{2020}  \at{{Convective heat
  transfer coefficient for a rod-like particle in a uniform flow}}.
  \jt{International Journal of Heat and Mass Transfer}  \bvol{147}.

\bibitem[Masliyah \& Epstein(1972)]{Masliyah1972}
{\sc \au{Masliyah, Jacob~H.} \& \au{Epstein, Norman}} \yr{1972} {Numerical
  solution of heat and mass transfer from spheroids in steady axisymmetric
  flow}.  \bt{In {\em Progress in Heat and Mass Transfer, Volume 6. Proceedings
  of the International Symposium on Two-Phase Systems\/} (ed. \ed{G.~Hetsroni
  \& S.~Sideman})},  \pg{pp. 613--632}.  \publ{Oxford: Pergamon}.

\bibitem[Masoud \& Stone(2019)]{Masoud2019}
{\sc \au{Masoud, Hassan} \& \au{Stone, Howard~A}} \yr{2019}  \at{{The
  reciprocal theorem in fluid dynamics and transport phenomena}} .

\bibitem[Michaelides(2003)]{Michaelides2003}
{\sc \au{Michaelides, Efstathios~E.}} \yr{2003}  \at{{Hydrodynamic force and
  heat/mass transfer from particles, bubbles, and drops - The freeman scholar
  lecture}}.  \jt{Journal of Fluids Engineering, Transactions of the ASME}
  \bvol{125}~(2),  \pg{209--238}.

\bibitem[Myerson(2002)]{HandbookCrystallisation2002}
{\sc \au{Myerson, Allan}} \yr{2002} {\em {Handbook of Industrial
  Crystallization}\/}.  \publ{Boston: Butterworth-Heinemann}.

\bibitem[Pahlow {\em et~al.\/}(1997)Pahlow, Riebesell \&
  Wolf-Gladrow]{Pahlow1997}
{\sc \au{Pahlow, Markus}, \au{Riebesell, Ulf} \& \au{Wolf-Gladrow, Dieter~A.}}
  \yr{1997}  \at{{Impact of cell shape and chain formation on nutrient
  acquisition by marine diatoms}}.  \jt{Limnology and Oceanography}
  \bvol{42}~(8),  \pg{1660--1672}.

\bibitem[Poe \& Acrivos(1976)]{Poe1976}
{\sc \au{Poe, G.G.} \& \au{Acrivos, A.}} \yr{1976}  \at{{Closed Streamline
  Flows Past Small Rotating Particles: Heat Transfer at High Peclet Numbers}}.
  \jt{Int. J. Multiphase Flow}  \bvol{2}~(4),  \pg{365--377}.

\bibitem[Pozrikidis(1997)]{Pozrikidis1997}
{\sc \au{Pozrikidis, C.}} \yr{1997}  \at{{Unsteady heat or mass transport from
  a suspended particle at low P{\'{e}}clet numbers}}.  \jt{Journal of Fluid
  Mechanics}  \bvol{334},  \pg{111--133}.

\bibitem[Sehlin(1969)]{Sehlin1969}
{\sc \au{Sehlin, R.C.}} \yr{1969}  \at{{Forced-Convection Heat and Mass
  Transfer at Large Peclet Numbers From an Axisymmetric Body in Laminar Flow:
  Prolate and Oblate Spheroids}}. PhD thesis, Carnegie-Mellon University,
  Pittsburgh, PA.

\bibitem[Seidensticker {\em et~al.\/}(2017)Seidensticker, Zarfl, Cirpka,
  Fellenberg \& Grathwohl]{Seidensticker2017}
{\sc \au{Seidensticker, Sven}, \au{Zarfl, Christiane}, \au{Cirpka, Olaf~A.},
  \au{Fellenberg, Greta} \& \au{Grathwohl, Peter}} \yr{2017}  \at{{Shift in
  Mass Transfer of Wastewater Contaminants from Microplastics in the Presence
  of Dissolved Substances}}.  \jt{Environmental Science and Technology}
  \bvol{51}~(21),  \pg{12254--12263}.

\bibitem[Sparrow {\em et~al.\/}(2004)Sparrow, Abraham \& Tong]{Sparrow2004}
{\sc \au{Sparrow, Ephraim~M.}, \au{Abraham, John~P.} \& \au{Tong, Jimmy~C.K.}}
  \yr{2004}  \at{{Archival correlations for average heat transfer coefficients
  for non-circular and circular cylinders and for spheres in cross-flow}}.
  \jt{International Journal of Heat and Mass Transfer}  \bvol{47}~(24),
  \pg{5285--5296}.

\bibitem[Suhrhoff \& Scholz-B{\"{o}}ttcher(2016)]{Suhrhoff2016}
{\sc \au{Suhrhoff, Tim~Jesper} \& \au{Scholz-B{\"{o}}ttcher, Barbara~M.}}
  \yr{2016}  \at{{Qualitative impact of salinity, UV radiation and turbulence
  on leaching of organic plastic additives from four common plastics — A lab
  experiment}}.  \jt{Marine Pollution Bulletin}  \bvol{102}~(1),  \pg{84--94}.

\bibitem[Szeri(1993)]{Szeri1993}
{\sc \au{Szeri, A.~J.}} \yr{1993}  \at{{Pattern formation in recirculating
  flows of suspensions of orientable particles}}.  \jt{Philosophical
  Transactions - Royal Society of London, A}  \bvol{345}~(1677),
  \pg{477--506}.

\end{thebibliography}

\end{document}